\documentclass[a4paper,fleqn,usenatbib]{mnras}

\usepackage{amsmath}	
\usepackage{txfonts}
\usepackage{mathrsfs}
\usepackage{breqn}

\usepackage[T1]{fontenc}
\usepackage{ae,aecompl}

\usepackage{graphicx}	
\usepackage{amssymb}	
\usepackage{multirow}
\usepackage{hyperref}
\hypersetup{
    colorlinks,
    citecolor=blue,
    filecolor=black,
    linkcolor=blue,
    urlcolor=black
}

\newcommand{\fesc}{\ifmmode{f_{\rm esc}}\else{$f_{\rm esc}$}\fi}
\newcommand{\fescs}{\ifmmode{f_{\rm esc}^\star}\else{$f_{\rm esc}^\star$}\fi}
\newcommand{\kms}{\ifmmode{{\;\rm km~s^{-1}}}\else{km~s$^{-1}$}\fi}
\newcommand{\fgas}{\ifmmode{{f_{\rm gas}}}\else{$f_{\rm gas}$}\fi}
\newcommand{\cubecm}{\ifmmode{{\rm cm^{-3}}}\else{cm$^{-3}$}\fi}
\newcommand{\ztwo}{\ifmmode{{\rm [Z_2/H]}}\else{[Z$_2$/H]}\fi}
\newcommand{\zthree}{\ifmmode{{\rm [Z_3/H]}}\else{[Z$_3$/H]}\fi}
\newcommand{\lsim}{\lower0.3em\hbox{$\,\buildrel <\over\sim\,$}}
\newcommand{\gsim}{\lower0.3em\hbox{$\,\buildrel >\over\sim\,$}}

\newcommand{\sfr}{\ifmmode{\textrm{M}_\odot \,\textrm{yr}^{-1} \,\textrm{Mpc}^{-3}}\else{M$_\odot$ yr$^{-1}$ Mpc$^{-3}$}\fi}
\newcommand{\hsfr}{\ifmmode{\textrm{M}_\odot\, \textrm{yr}^{-1}}\else{M$_\odot$ yr$^{-1}$}\fi}

\newcommand{\eavg}{\ifmmode{\langle E_\gamma \rangle}\else{$\langle E_\gamma \rangle$}\fi}

\newcommand{\enzo}{{\sc enzo}}

\newcommand{\Ms}{\ifmmode{M_\odot}\else{$M_\odot$}\fi}
\newcommand{\vrms}{\ifmmode{v_{\rm rms}}\else{$v_{\rm rms}$}\fi}

\newcommand{\tvir}{\ifmmode{T_{\rm{vir}}}\else{$T_{\rm{vir}}$}\fi}
\newcommand{\mvir}{\ifmmode{M_{\rm{vir}}}\else{$M_{\rm{vir}}$}\fi}
\newcommand{\rvir}{\ifmmode{r_{\rm{vir}}}\else{$r_{\rm{vir}}$}\fi}

\newcommand{\jj}{\ifmmode{J_{21}}\else{$J_{21}$}\fi}
\newcommand{\flw}{\ifmmode{F_{LW}}\else{$F_{LW}$}\fi}
\newcommand{\kph}{\ifmmode{k_{\rm ph}}\else{$k_{\rm ph}$}\fi}

\newcommand{\zsun}{\ifmmode{\rm\,Z_\odot}\else{$\rm\,Z_\odot$}\fi}

\newcommand{\nhi}{\ifmmode{N_{\rm HI}}\else{$N_{\rm HI}$}\fi}

\def\eps@scaling{1.0}%
\newcommand\epsscale[1]{\gdef\eps@scaling{#1}}%
\newcommand\plotone[1]{%
 \centering 
 \leavevmode 
 \includegraphics[width={\eps@scaling\columnwidth}]{#1}%
}%
\newcommand\plottwo[2]{%
 \centering 
 \includegraphics[width={\eps@scaling\columnwidth}]{#1}%
 \hfil 
 \includegraphics[width={\eps@scaling\columnwidth}]{#2}%
}%

\title[Emission Lines and Population III Stars]{First Light II: Emission Line Extinction, Population III Stars, and X-ray Binaries}

\author[K. S. S. Barrow et al.]{Kirk S. S. Barrow$^{1,2}$\thanks{e-mail:
    kssbarrow@gmail.com}, John H. Wise$^{1}$, Aycin Aykutalp$^{1,3}$, Brian W. O'Shea$^4$, \newauthor  Michael L. Norman$^5$, and Hao Xu$^{5,6}$\\
  $^{1}$ Center for Relativistic Astrophysics, Georgia Institute of
  Technology, 837 State Street, Atlanta, GA
  30332, USA\\
  $^2$ Jet Propulsion Laboratory, National Aeronautics and Space Administration, Pasadena, CA 91109, USA\\
  $^3$ Los Alamos National Laboratory, Los Alamos, NM 87545, USA\\
  $^4$ 
Department of Computational Mathematics, Science and Engineering, Department of Physics and Astronomy, \\
  \ \ \ and National Superconducting Cyclotron Laboratory, Michigan State University, East Lansing, MI 48824, USA\\
  $^{5}$ CASS, University of California, San Diego, 9500 Gilman Drive, La Jolla, CA 92093, USA\\
  $^6$ IBM, New Orchard Road, Armonk, NY 10504, USA
}
\pagerange{\pageref{firstpage}--\pageref{lastpage}} \pubyear{2017}

\date{Accepted 2017 November 9; Received 2017 October 24; in original form 2017 September 11
}

\begin{document}
\label{firstpage}
\pagerange{\pageref{firstpage}--\pageref{lastpage}}
\maketitle

\begin{abstract}
  
We produce synthetic spectra and observations for metal-free stellar populations and high mass X-ray binaries in the Renaissance Simulations at a redshift of 15. We extend our methodology from the first paper in the series by modelling the production and extinction of emission lines throughout a dusty and metal-enriched interstellar and circum-galactic media extracted from the simulation, using a Monte Carlo calculation. To capture the impact of high-energy photons, we include all frequencies from hard X-ray to far infrared with enough frequency resolution to discern line emission and absorption profiles. The most common lines in our sample in order of their rate of occurrence are Ly$\alpha$, the C IV $\lambda\lambda1548,1551$ doublet, H-$\alpha$, and the Ca II $\lambda\lambda\lambda8498,8542,8662$ triplet. The best scenario for a direct observation of a metal-free stellar population is a merger between two Population III galaxies. In mergers between metal-enriched and metal-free stellar populations, some characteristics may be inferred indirectly. Single Population III galaxies are too dim to be observed photometrically at $z = 15$. Ly$\alpha$ emission is discernible by $JWST$ as an increase in $\rm{J_{200w} - J_{277w}}$ colour off the intrinsic stellar tracks. Observations of metal-free stars will be difficult, though not impossible, with the next generation of space telescopes.

\end{abstract}


\section{Introduction and Background}

The advent of large and high-resolution cosmological simulations such as the Renaissance Simulations \citep{2015ApJ...807L..12O,2016ApJ...833...84X} provide an opportunity to glean observables from theoretical and numerically-deduced phenomena. However because radiative transfer is computationally expensive inside a full simulation, post-processing is usually required to better extrapolate the fine features of the spectral energy distribution related emission lines and dust extinction.

Therefore the contemporary frontier of synthetic spectrometry and photometry lies in the sophistication and physical accuracy of post-processing techniques. As the second paper in the series chronicling the methodological development of the {\sc Caius} pipeline, this work seeks to invest more computational power in the modelling of absorption, scattering, and emission of photons and explores the impact of high-energy photons from sources unique to the early universe.

\subsection{Population III stars and X-ray Binaries}

Prior to the formation of the first generation of stars, the primordial medium was essentially metal-free. The lack of metal cooling theoretically increases the Jeans mass of molecular cloud and allows for the possibility for the formation of more massive stars than the observed initial mass function (IMF) of stars in the local universe. However the primordial IMF is limited at the upper range by the effect of radiation pressure, the nature of the protostar, the effective sound speed, and the propensity of the medium to clump \citep[see reviews by][]{2007ARA&A..45..565M,2015ComAC...2....3G}. These early ``Population III" stars were luminous and hot for their masses and thus contributed to the ionization and heating of the surrounding medium.

Massive Population III stars are likely to end in neutron stars and black holes after a short mass-dependent lifespan \citep{2002A&A...382...28S}. When part of binary systems, compact remnants may thereupon accrete material from a longer-lived stellar partner, converting potential energy into high-energy photons in their accretion disks. In this scenario, termed an X-ray binary, the flux of X-rays and UV photons ionize metals in the interstellar medium (ISM) to multiply-ionized states like {\sc C IV}. Indeed, {\sc C IV} lines have been confirmed in the spectrum of analogous low-redshift systems of low mass ($M_\star < M_\rm{bh}$) X-ray binaries \citep[e.g.][]{2005ApJ...634L.105D}, and high mass X-ray binaries \citep[HMXB; e.g.][]{2007ASPC..367..459I}.

\subsection{Emission Lines}

While there has been progress on the modelling of emission lines and dust \citep{2013ApJ...777...39Z,2014ApJ...782...32C,2016MNRAS.460.3170W,2017MNRAS.469.4863B}, the work of solving photoionization is usually left to routines or analysis run on isolated scenarios rather than within a cosmological context. This work endeavours to simulate photochemistry as part of an extinction and emission routine that appreciates a fully three-dimensional arrangement of dust, gas, and stars. In Sec. \ref{sec:meth}, we propose a methodology for generating and propagating the intrinsic spectrum of metal-enriched stars, metal-free Population III stars, and HMXBs through gas and dust to produce resultant galactic continuum and emission lines. In Sec. \ref{sec:resul}, diagnostics of observables from our treatment are presented including emission line strengths and photometry for the forthcoming James Webb Space Telescope (JWST). Finally, the observational and physical implications of our results are discussed in Sec. \ref{sec:diss} and summarized in Sec. \ref{sec:con}.

\section{Methods}

\label{sec:meth}

We use the ``rare-peak" zoom-in region of the Renaissance Simulations \citep{2015ApJ...807L..12O,2016ApJ...833...84X}, which are performed using the hydrodynamic adaptive-mesh refinement (AMR) code \enzo{} with radiative transfer \citep{2011MNRAS.414.3458W} and focus on the first generations of stars and galaxies early in the Epoch of Reionization. The simulations are run using $\Omega_M = 0.266$, $\Omega_{\Lambda} = 0.734 $, $\Omega_b = 0.0449$, $h = 0.71$, $\sigma_8 = 0.8344$, and $n = 0.9624$ from the 7-year $Wilkinson\ Microwave\ Anisotropy\ Probe$ results \citep[WMAP;][]{2011ApJS..192...16L}, achieving an effective dark matter resolution of $2.9 \times 10^4\ \rm{M_\odot}$ at $z = 15$ and a spatial resolution of 19 comoving parsecs. We identified 1654 galaxies containing stellar clusters within the ``rare-peak" simulation at $z= 15$ using dark matter halo-finding code {\sc Rockstar} \citep{2013ApJ...762..109B}. Of these galaxies, we focus our analysis on 146 that contain metal-free stellar populations.

\subsection{Stellar Spectra}

We use the Flexible Stellar Population Synthesis code \citep[FSPS;][]{2010ApJ...712..833C} to generate source spectra of metal-enriched stellar clusters from the age and metallicities of particles in the simulation. Both FSPS and the simulation routines treat stellar clusters as probabilistic distributions of stars and treat luminosity as a linear function of mass. We exploit this similarity to assign spectral energy distributions (SEDs) to stellar cluster particles for radiative transfer post-processing in a manner consistent with the assumptions used to produce those particles. Using quantities calculated in the simulation, we assign each particle a metallicity isochrome and allow FSPS to interpolate an SED based on the particle's age before weighting our result by the mass of the cluster.

\begin{figure*}
\begin{center}
\includegraphics[scale=.36]{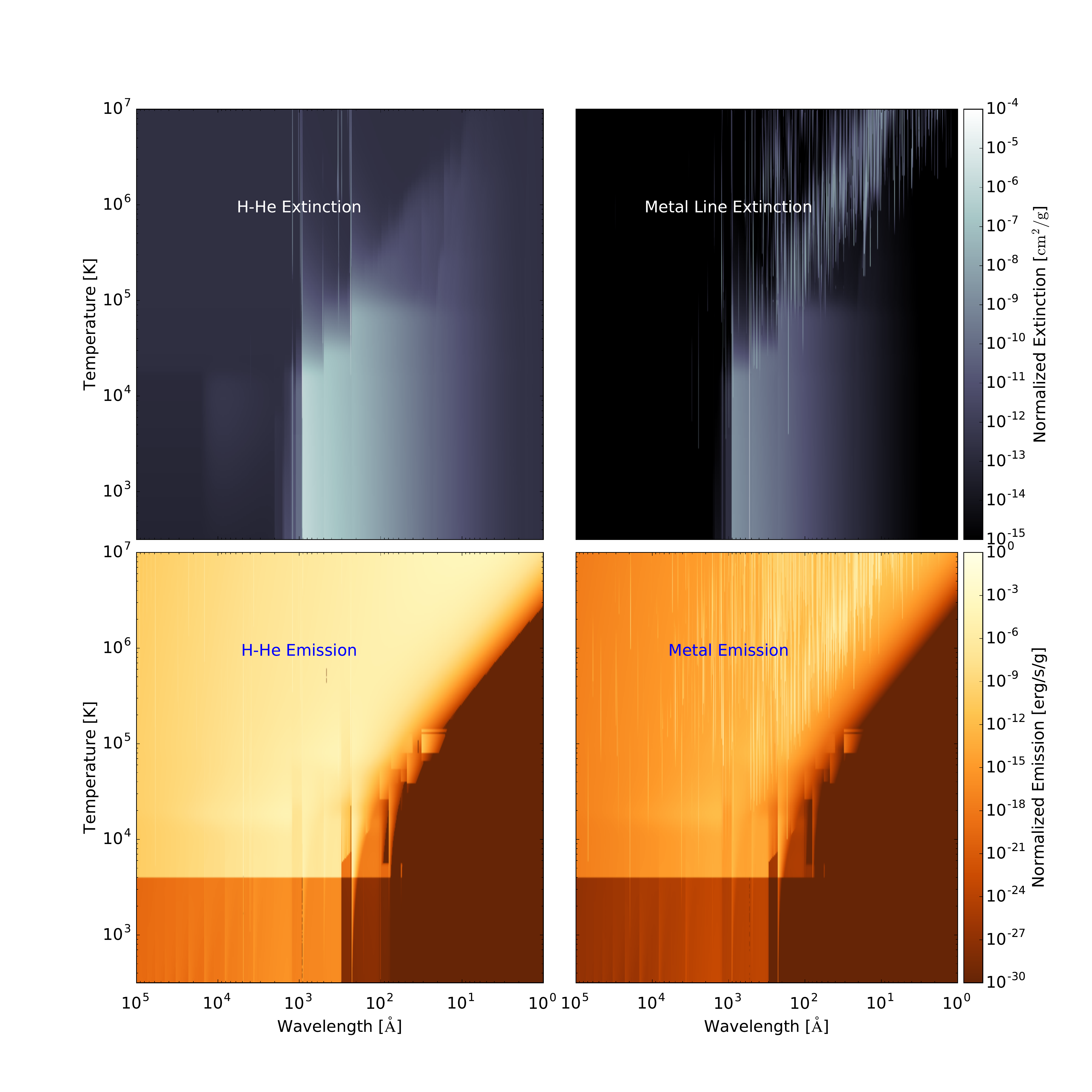}
\caption{Top Row: Total extinction profiles for gas and metal emission lines and a gas absorption continuum as a function of wavelength and temperature. Bottom Row: Corresponding emission profiles assuming thermodynamic equilibrium.}
\label{fig:plotfour}
\end{center}

\end{figure*}

For metal-free stars, the Renaissance Simulations generate stellar particles corresponding to individual stars rather than clusters in the simulation by randomly sampling the distribution

\begin{equation}
  f(\log M)dM = M^{-x} \exp\left[-\left(\frac{M_c}{M}\right)^{\beta} \right] dM,
  \label{eq:IMF}
\end{equation}  
when the appropriate environmental conditions are achieved \citep{2017MNRAS.469.4863B}. This form obeys the \citet{1955ApJ...121..161S} distribution above a characteristic mass, $M_c$, and has an exponential cut off at low mass \citep{2003PASP..115..763C,2012MNRAS.427..311W}. We limit Population III star particle mass to the range $1\ {\rm M}_\odot$ to $300\ {\rm M}_\odot$ and use $x = 1.3$, $\beta = 1.6$, and $M_c = 40\ {\rm M}_\odot$. Generally, the peak of the distribution is given by

\begin{equation}
M_{\rm{peak}} =  M_c \left(\frac{\beta}{x}\right)^{1/\beta},
\end{equation}
which corresponds to $\sim 46 \ {\rm M}_\odot$ in our simulation after accounting for the mass  cut-offs. As noted by \citet{2011ApJ...740...13Z} in the description of the {\sc Yggdrasil} metal-free stellar SED calculator, the exact form of the initial mass function (IMF) and therefore the spectra of Population III stars is a matter of great uncertainty. {\sc Yggdrasil} offers three prescriptions with three different IMFs. The PopIII.1 tool assumes a \citet{2002A&A...382...28S} power law with slope $\alpha = -2.35$ and stellar masses from 50 to 500 ${\rm M}_\odot$ and models the most antecedent generation of stars prior to any radiative feedback and binary formation. The PopIII.2 model assumes a log-normal distribution with a characteristic mass of 10 ${\rm M}_\odot$ and models an IMF that includes stars from 1 to 500 ${\rm M}_\odot$ assuming primordial abundances but some impact from prior star formation and radiative processes \citep{2008AIPC..990D..13O}. The third model assumes a metal-enriched IMF \citep{2001MNRAS.322..231K} with a piece-wise power law for different mass regimes.

For this work, we determined the PopIII.1 model to be too top heavy for the lower end of our mass range and the Kroupa model to be inconsistent with the goal of differentiating metal-enriched and metal-free stellar populations so we generate our SED using the PopIII.2 model for stars smaller than 55 ${\rm M}_\odot$ and the PopIII.1 model for larger stars. We generate and match SEDs to the age and mass of the simulation star particles assuming an instantaneous burst. Since we treat the effect of extinction separately in our calculations, we do not include a covering fraction to produce the SED. We note that there are some discrepancies between flux of ionizing radiation modelled in the simulation and the flux calculated from {\sc Yggdrasil} due to our practice of using an average IMF to model individual stars in post-processing. We expect this to manifest as inconsistencies between the flux of ionizing radiation from the metal-free SEDs and the size of ionized regions in the simulation so we hereafter give priority to the values of electron fraction and temperature from the simulation in our radiative transfer analysis.

\subsection{Emission Line Extinction}

\begin{figure*}
\begin{center}
\includegraphics[scale=.34]{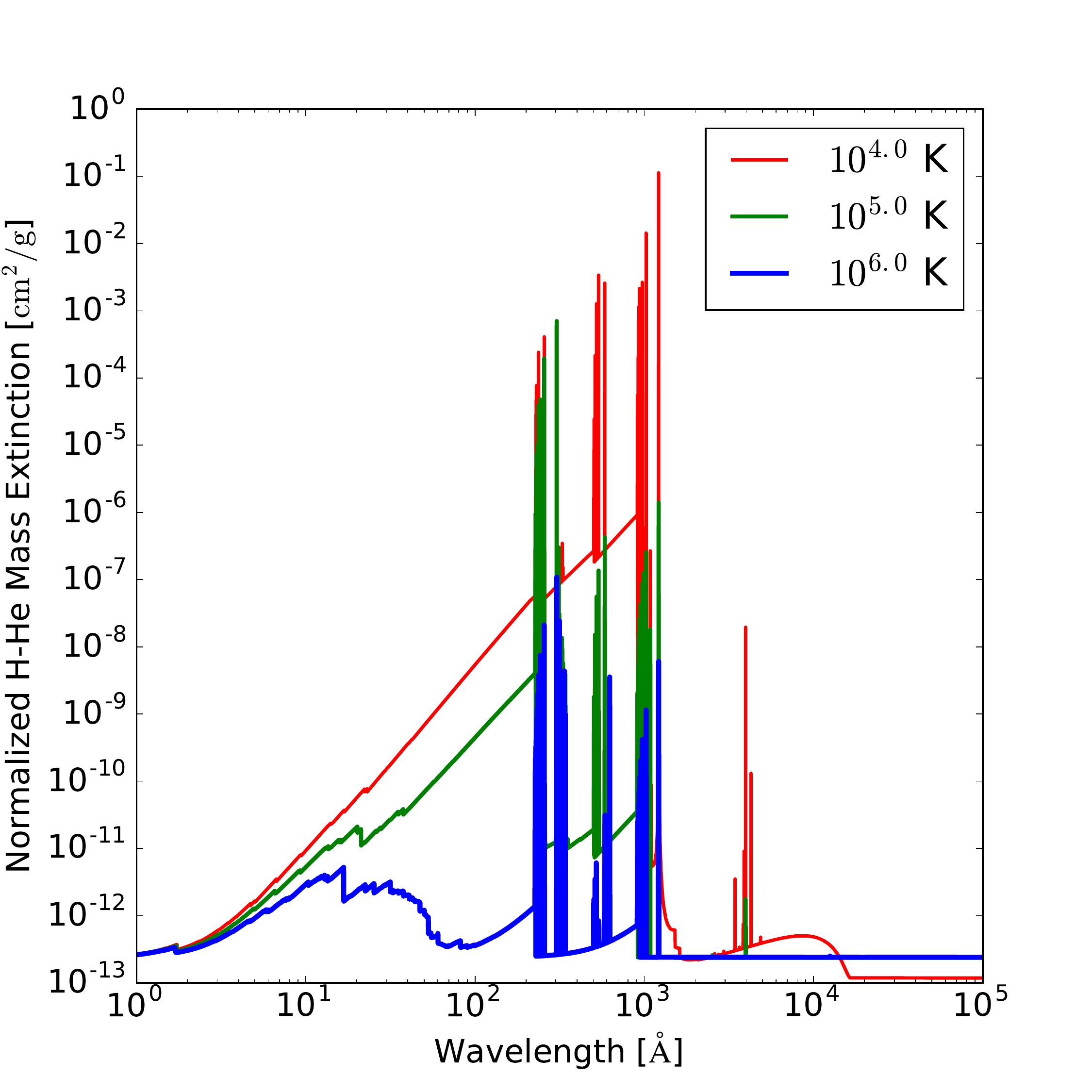} 
\includegraphics[scale=.34]{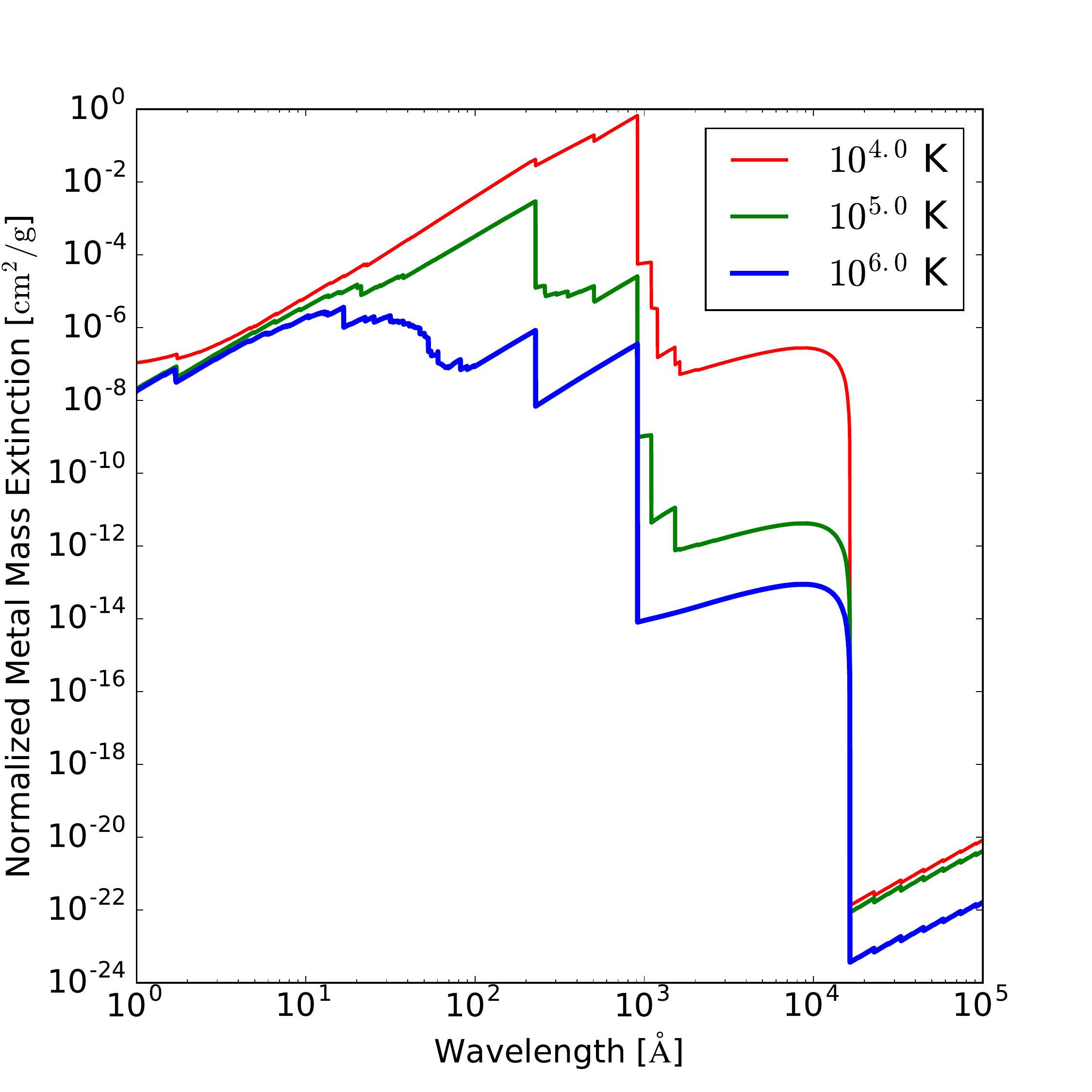}
\caption{Isothermal cross-sections of extinction profiles. Left: Mass extinction coefficient for hydrogen and helium continuum and lines. Right: Mass extinction coefficient for metal lines with a hydrogen and helium continuum. Both plots are normalized by their maximum value in Fig. \ref{fig:plotfour} and are plotted with thin, medium, and thick lines for gas of temperatures $10^4$,$10^5$, and  $10^6$ respectively.}
\label{fig:emcross}
\end{center}

\end{figure*}

The antecedent treatment of emission lines from \citet{2017MNRAS.469.4863B} used the photo-ionization solver {\sc Cloudy} \citep{2013RMxAA..49..137F} to determine emission line strength in simulation AMR cells containing stellar populations.  Line emissions were therefore limited to a small fraction of the interstellar medium (ISM) and almost none of the circum-galacitic medium (CGM). A separate Monte Carlo gas and dust extinction calculation of the continuum using {\sc Hyperion} \citep{2011A&A...536A..79R} was used to attenuate the lines in conjunction with an empirical correction from \citet{2015MNRAS.454..269P}. In this work, we more thoroughly examine the use of {\sc Cloudy} and {\sc Hyperion} as tools for emission line extinction and line transfers in arbitrary stellar, dust, and gas arrangements while eschewing the use of empirical attenuation models.

In a similar manner to our prior method, we use gas densities and metallicities from the cosmological simulation and stellar spectra to stage a {\sc Cloudy} calculation for AMR cells containing interior stellar populations. For cells with more than one stellar particle, which may include combinations of metal-enriched clusters and metal-free stars, spectra are summed into a single source for the calculation. In the simulation, stellar particles are formed within a single cell by design, but may move between cells after their formation. As a result, the highest refinement level within a halo usually contains one, or at most, a few stellar particles. However, the cells contain too little medium to calculate the photochemistry in the ISM after the tenth level of refinement. Therefore we attempt to balance this with our desire to limit the use of particle SED summing by allowing up to the ninth level of the simulation AMR grid. We run {\sc Cloudy} until the electron fraction matches the value from the corresponding cell rather than allow the calculation to come to a thermal equilibrium and apply Doppler broadening to the lines by using the temperature of the cell and the mass of the emitting molecule. We take enough samples of the spectra to produce a discernible Gaussian distribution of most lines, which increases the computational load of our pipeline considerably when compared to our prior investigation. The line profiles are then redistributed back to the intrinsic spectra of each particle proportionally to their fraction of the total luminosity of the summed source. The result is usually a relatively small addition to the spectra from diffuse emission, but we include this calculation in our method to capture any unique photochemical interactions in regions with high flux from a local source.

In addition to the lines added to the intrinsic stellar spectra, we model emission and absorption more generally throughout the interstellar and circumstellar medium. To account for a chemically inhomogeneous interstellar medium, we treat our halos as similarly inhomogeneous distributions of metallicity by using the precise emission line wavelengths and source molecule to segregate line opacities and emissions generated by non-metals from those generated by metals. We calculate extinction, albedo, and emission for 400 equally log-spaced temperatures between $10^{2.5}$ and $10^{7}$ K using a flat spectrum in thermodynamic equilibrium and constant density and metallicity. For simplicity, we assume solar abundance patterns \citep{2009ARA&A..47..481A} when the metallicity of the gas exceeds $10^{-6}\ {\rm Z}_\odot$ and turn off the presence of metals entirely below that value.  We note that, for example, carbon line absorption from high-redshift quasars demonstrate a nearly constant carbon column density throughout reionization \citep[e.g.][]{2006ApJ...653..977S}. This implies that early enrichment biased away from the products of Type Ia supernovae and that the relative contribution from elements like carbon and silicon may be understated in this work.

Extinction of lines is calculated by generating a high-resolution frequency-dependent line opacity map with {\sc Cloudy} for frequencies corresponding to emission lines and adding the result to the absorption profile of the continuum. 
As shown in Fig. \ref{fig:plotfour}, this allows us to create a profile for extinction that scales with metal density and one that scales with non-metal density. Notably, the emission profiles show a clear delineation between temperatures that represent thermally ionized hydrogen and neutral species at around 4000 K which corresponds roughly to 50\% ionization of hydrogen according to the Saha equation. Both the emission and extinction profiles include the bremsstrahlung effect. For the metal profiles, we include the continuum extinction and emission due to hydrogen and helium by necessity to ensure that each photon has some extinction, but reason that this only accounts for a negligible over-accounting of extinction by gas. Fig. \ref{fig:emcross} shows how Gaussian broadening results in overlapping absorption profiles for species at low wavelengths, allowing metals to absorb X-ray emission efficiently. Additionally we use a third profile for dust opacities using \citet{2003ARA&A..41..241D} ($\rm{R_v} = 2.1$) that we scale as a fraction of the metal density.

Gas, metal, and dust extinction and emission profiles are used to perform a radiative transfer calculation with {\sc Hyperion} by propagating $5 \times 10^8$ photons at $5 \times 10^4$ log-spaced wavelengths from 10 to 50000 \AA (rest). These values are chosen such that they have sufficient resolution to capture line profiles of up to third period elements and calculate rest-frame radiative interactions from infrared through X-ray photon energies. While final emission profiles interpolate all values in the temperature range, we are forced to average extinction profiles into ten bands for the Monte Carlo algorithm. As shown in Fig. \ref{fig:plotfour}, extinction for individual species usually extends across a broad range of temperatures and since the medium within each cell would likely exhibit a range of temperatures at higher resolution, our use of averaged bands is still representative of the physical analogues to our simulations without losing too many features in the profile.  Compared to the antecedent method, the inclusion of opacities and emission profiles allows for the treatment of line transfer of photons from external sources, anisotropic absorption and emission lines, and calculation for emission line strength from H~{\sc II} regions larger than the size of a cell. 

We estimate the specific radiative power of baryons in a cell by using the equation

\begin{equation}
\mathscr{E} = 4 \sigma T^4 \frac{\int^{\nu_{\rm{max}}}_{\nu_{\rm{min}}} (\kappa_\nu / \rho)  B_\nu (T,\nu) d\nu}{\int^{\nu_{\rm{max}}}_{\nu_{\rm{min}}} B_\nu (T,\nu) d\nu},
\end{equation}
in agreement with \citet{1999A&A...344..282L} where $B_\nu (T)$ is the Planck spectral radiance distribution given by

\begin{equation}
B_\nu (T,\nu) = \frac{2h \nu^3}{c^2} \frac{1}{e^{h \nu/ k T}-1}.
\end{equation}
The quantity $\kappa_\nu/\rho$ is the mass absorption coefficient formed by dividing the absorption opacity by the density of the medium. The other variables have their standard physical definitions. The specific power $\mathscr{E}$, determined in units of $\rm{erg\ s^{-1} g^{-1}}$, sets the emitted radiative power per unit mass.

\begin{figure}
\begin{center}
\includegraphics[scale=.24]{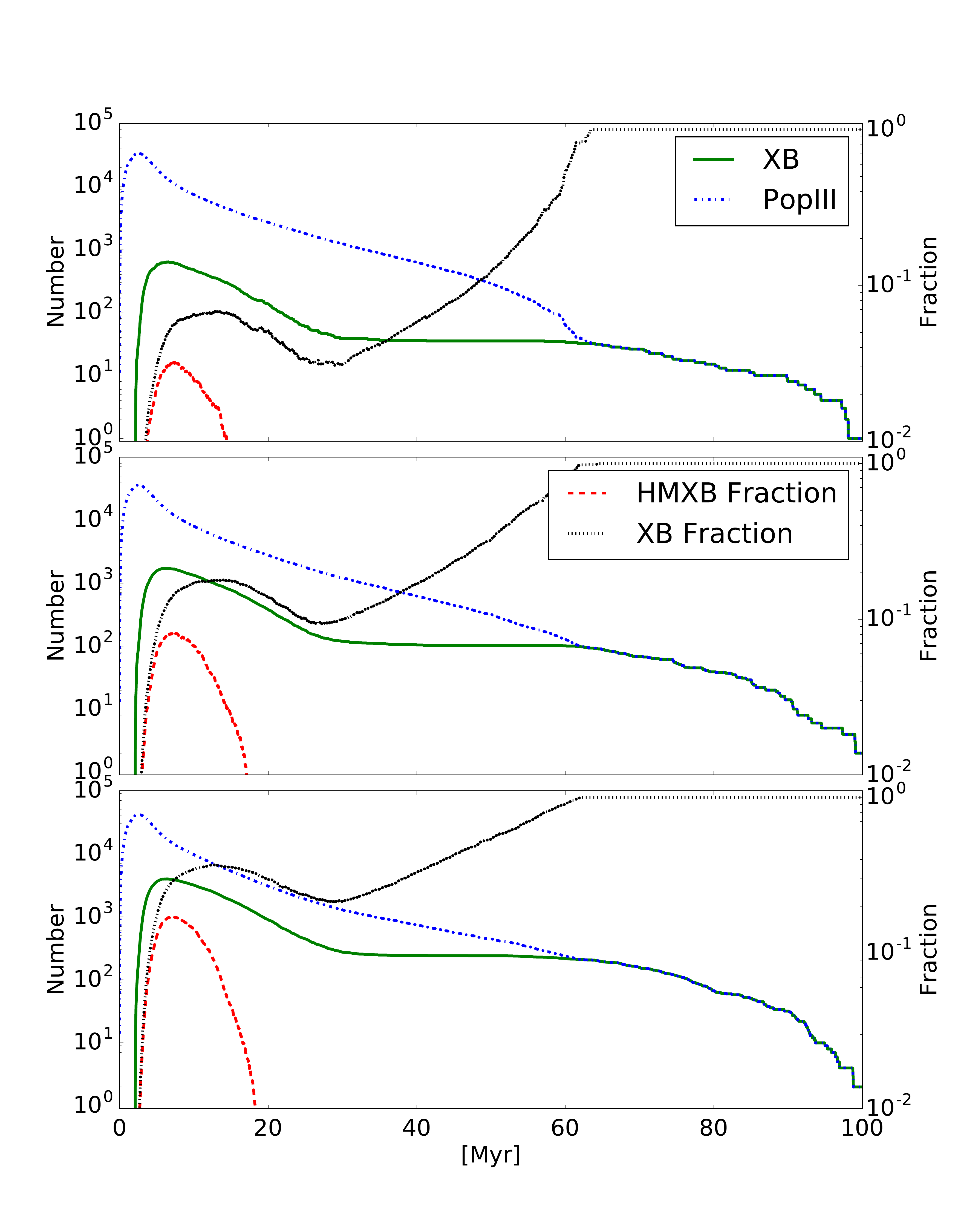}
\caption{Number of Population III stars (green), number of X-ray binaries (blue), fraction of X-ray binaries (black), and fraction of high mass X-ray binaries (red) for a burst of star formation assuming that 10\%(top), 25\%(middle), and 50\%(bottom) of the stars form initially in close binaries and accrete at the Eddington limit. The persistence of an individual particle is largely a function of sampling of the IMF, but the number and fraction of HMXBs is related to the close binary fraction.}
\label{fig:mbinary}
\end{center}

\end{figure}

We find that some lines and scenarios with high scattering opacities prove to be infeasible to model using a Monte Carlo method so we rescale the scattering and absorption coefficients to produce a physically similar phenomenon with fewer scatterings. We reason that as long as the quantity $L \times \kappa_\nu$ is a constant where $L$ is the path length of an individual photon, the probability of absorption remains fixed. We also reason that as long as a photon scatters at least once within a cell, the final direction of the photon is indistinguishable from a scenario where it scatters many times. Therefore, we note that the dispersion of the radius of a three-dimensional random walk is given by $\langle R^2 \rangle = N \lambda^2$ where $R$ is the radius from the starting position, $N$ is the number of scatterings, and $\lambda = 1/\kappa_s$ is the mean free path. The path length is simply $L = N \lambda$ and the average number scatterings before crossing a cell is $N_1 = R^2 \kappa_{s,1}^2$. For a chosen maximum feasible number of scatterings $N_2$, we determine the corresponding scattering coefficient to be 

\begin{equation}
\kappa_{s,2} = \frac{\sqrt{N_2}}{R}.
\end{equation} 
Likewise because we take the path length times the absorption coefficient to be constant

\begin{equation}
\frac{L_1}{L_2} = \frac{\kappa_{\nu,2}}{\kappa_{\nu,1}} = \frac{\kappa_{\rm{s},1}}{\kappa_{\rm{s},2}},
\end{equation}
and we determine the corresponding absorption coefficient to be

\begin{equation}
\kappa_{\nu,2} = \frac{R \kappa_{s,1} \kappa_{\nu,1}}{\sqrt{N_2}}.
\end{equation}
Therefore for a given cell width, an opacity distribution, and a predetermined number of scatterings, we can recreate a physically similar scenario to the true opacities in optically thick regions. We choose $N_2$ to be 1000 to ensure that scatterings occur with an appropriately high resolution within a cell and only apply this correction to situations where $N_1 > N_2$. As an example, $N_2$ = 1000 corresponds to photons with a Lyman-$\alpha$ scattering cross-section of $10^{-16}\ \rm{cm^2}$ in a 10 parsec box and a hydrogen density of $\sim 1.7 \times 10^{-26}\ \rm{ g\ cm^{-3}}$.

\subsection{High Mass X-Ray Binaries}
\label{sec:HMXB}

\begin{figure}
\begin{center}
\includegraphics[scale=.37]{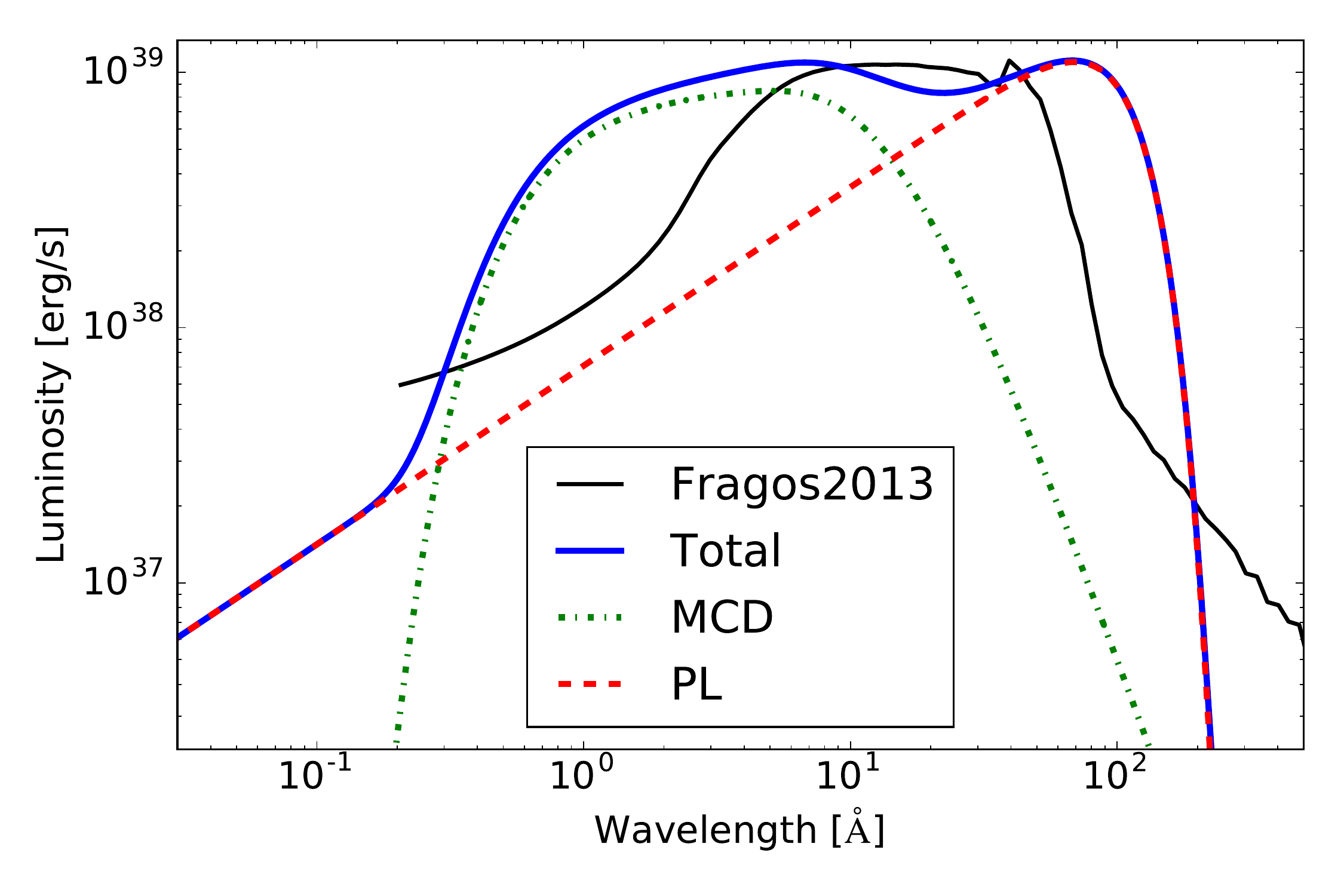}
\caption{Multi-color disk (MCD) and power law (PL) components of a SED from a 40 ${\rm M}_\odot$ black hole accreting at the Eddington Limit. Absorption at wavelengths above 200 $\AA$ attenuate the power law considerably. A re-normalized $z=15.3$ \citet{2013ApJ...776L..31F} intrinsic HMXB spectra is shown for comparison.}
\label{fig:black3}
\end{center}

\end{figure}

Armed with theoretical spectra of Population III stars and a generalized emission line routine, we extend our investigation to the impact of high mass X-ray binaries. 

\subsubsection{Luminous X-Ray Binary Fraction Simulation}

Self-consistent formation of binary metal-free stars are outside of the scope of the cosmological simulation so we attempt to set reasonable bounds to parameters associated with their population using a semi-analytical treatment. We simulate the life cycle of a burst of Population III stars with masses sampled from the Population III IMF (Equation \ref{eq:IMF}), mass-dependent lifetimes from \citep{2002A&A...382...28S}, and an IMF-dependent stellar endpoints \citep{2003ApJ...591..288H} including the possibility that no remnant is left in the case of a pair-instability supernovae. To even out statistical noise, we simulate a metal-free starburst of 36,050 systems assuming scenarios where half, 25\%, and 10\% of systems are formed as ``close binaries." That is, the stars are close enough that should the shortest-lived member form a black hole or neutron star at the end of its life, the longer-lived member will accrete its mass onto the remnant and thus form an X-ray binary. The number of systems comes from the integration of a star formation rate that peaks in at one million years and peters out exponentially over the next four million years. We note that HMXB populations synthesis models in \citep{2013ApJ...764...41F} describe an inverse relationship between metallicity and HMXB number as weaker stellar winds in low-metallicity gas promotes the production of larger remnants and less angular momentum loss in tight binary orbits, which implies a higher fraction of HMXB than observed at lower redshift.

We further assume \citet{1926ics..book.....E} mass accretion rates for luminous compact objects and recalculate the lifetime and remnant type of stars as they lose mass. This assumption places a lower bound on the duration of HMXBs in the calculation. These scenarios are presented in Fig. \ref{fig:mbinary} which shows that the persistence of X-ray binary systems is largely a function of lucky sampling of the IMF. The maximum number of X-ray binaries in the 50\% calculation is $\sim 6.5$ times as many as in the 10\% calculation in this example, but this too is subject to the whims of random sampling for any individual halo. For context, target galaxies in the $Chandra$ Deep Field South survey have been shown to emit an X-ray luminosity (2-10 keV) from HMXBs of $\sim 3 \times 10^{39}\ \rm{erg\ s^{-1}}$ or about two continuous 40 $\rm{M_\odot}$ HMXB per $\rm{M_\odot\ yr^{-1}}$ star formation rate \citep{2016ApJ...825....7L} which grows to $\sim 3 \times 10^{40}\ \rm{erg\ s^{-1}}$ at $z=10$ \citep{2017ApJ...840...39M}. Our Population III bursts typically have 25 Myr averaged star formation rates that peak at $\sim 0.01\ \rm{M_\odot\ yr^{-1}}$, which implies that our average halo would have about one HMXB. The halo with the highest peak star formation rate corresponds to approximately 15 continuous HMXBs, but this is exaggerated by the top-heavy IMF of metal-poor stars.

For this study, we are less interested in the precise global fraction of HMXBs and more interested in whether they plausibly exist and therefore warrant study as a possible source for high energy photons in the SED of galaxies with metal-free stars. From our rough calculation, we can conclude that the presence of HMXBs are possible in any halo that once contained two or more metal-free stars from a few million years after the initial burst until about 17 Myr. Since our sample of halos focus on galaxies with mixed populations of metal-free and metal-enriched stars, we reason that most of our galaxies are subject to the region of the predicted HMXB distribution corresponding to 4-17 Myr after the starburst and indeed most halos have small populations of lower-mass, longer living metal-free stars. For that scenario, we convert as many as two of the Population III star particles into HMXBs. If the maximum age of the metal-free star particles in the halo is less than 2 Myr, we do not convert any of them into HMXBs because those stellar systems are too young to contain a compact object. We note that HMXBs are possible in metal-enriched stellar populations, but with lower frequency due to the propensity of metal-enriched gas to fragment and form less massive stars and fewer compact objects. We therefore do not include metal-enriched HMXBs in our study.

\subsubsection{X-Ray Binary Spectra}

\begin{figure}
\begin{center}
\includegraphics[scale=.35]{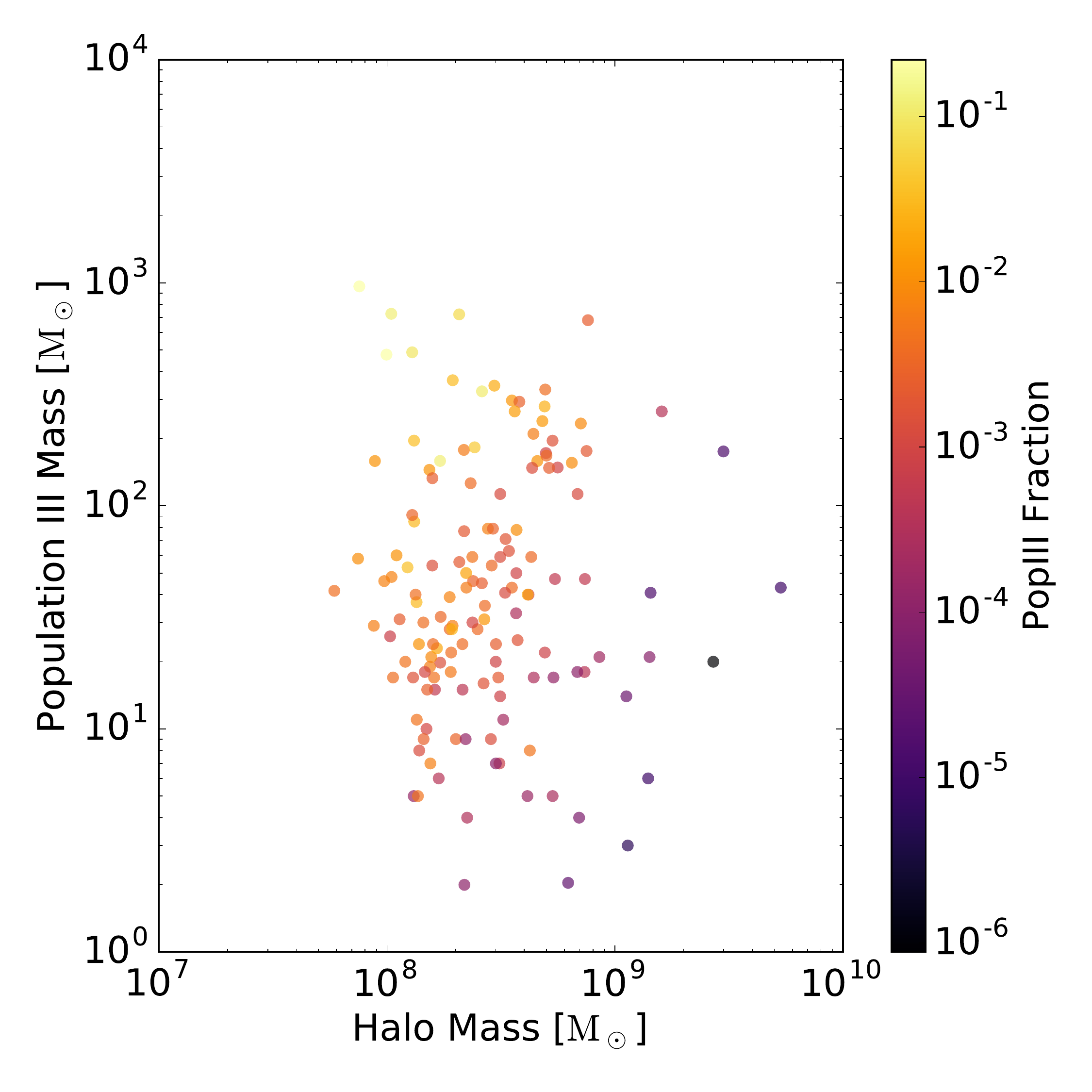}
\caption{Plot of total metal-free stellar mass versus total halo mass tinted by metal-free stellar fraction of the total stellar mass.}
\label{fig:stellarplt}
\end{center}

\end{figure}

For simplicity and because of Pop III IMF uncertainty, we assume a black hole mass equal to the simulation characteristic mass of 40  ${\rm M}_\odot$ to calculate its spectrum. We assume a radiative efficiency of 0.1 and that emission is equally distributed between a multi-color accretion disk and a power law of the form $\dot{E} \propto E^{-1.7}$ in units of eV.

For the multi-color disk we use the temperature profile from \citet{2003ApJ...597..780E} given by 

\begin{equation}
T_{\rm{eff}} = \left[\frac{3GM\dot{
M}}{8 \pi \sigma_{\rm{T}} r^3} \left(1-\sqrt{\frac{r_{\rm{in}}}{r}}\right)\frac{r_{\rm{in}}}{r} \right]^{1/4},
\end{equation}
where $\sigma_{\rm{T}}$ is the Thompson scattering cross section and the innermost radius, $r_{\rm{in}}$, is set to six gravitational radii. We also apply the correction $T_{\rm{col}} = 1.7\ T_{\rm{eff}}$ due to the Comptonization of the disk and calculate color temperatures out to 5000 gravitational radii. A black body distribution is calculated for each temperature and weighted by the factor $2 \pi r \Delta r$. The resulting distribution is finally normalized to half the Eddington luminosity, which is given by

\begin{equation}
L_{\rm{edd}} = \frac{4 \pi G M m_{\rm{p}} c}{\sigma_{\rm{T}}},
\end{equation}
where $m_{\rm{p}}$ is the mass of a proton. We also apply hydrogen and helium absorption to the power law assuming primordial abundances and distribute the other half of the black hole's luminosity to the absorbed result. For absorption, we assume a neutral hydrogen column density of log$[N_{HI}/\rm{cm^{-2}}]=20$ due to the accumulation of material in the vicinity of the accretion disk as the star is quickly disrupted and consumed. The assumption of sub-grid neutral hydrogen absorption does not strongly affect the amount of ionizing radiation fed through the medium in most halos due to the presence of strong flux at those wavelengths from Population III stars. The resulting intrinsic spectra for a 40 ${\rm M}_\odot$ black hole is shown in Fig. \ref{fig:black3} absent the spectra for the binary star.  A \citet{2013ApJ...776L..31F} intrinsic HMXB spectra calculated from a considered stellar population synthesis model for metal-enriched X-ray binaries is shown to have roughly similar features.

\subsection{Spectral Energy Distribution Analysis}

The filtering and imaging routines are effectively the same as those discussed in \citet{2017MNRAS.469.4863B}. To summarize, we calculate flux using $JWST$ and $HST$ filter throughputs to integrate the processed SED after applying cosmological corrections as a function of redshift. Images are created by summing photons intersecting a distant plane using {\sc Hyperion} and applying noise, Gaussian blur, and the telescope's resolution when processed through a telescope prescription.

For bolometric flux, the equations are

\begin{equation}
d_{\rm{L}} \ = \frac{c (1+z)}{H_{\rm{0}}}\int_0^z \frac{dz'}{\sqrt{\Omega_{\rm{M,0}} (1+z')^3 + \Omega_{\rm{\Lambda,0}}}}
\end{equation}

\begin{equation}
f(\nu_{\rm{0}}) =  \frac{1}{4 \pi d_{\rm{L}}^2} \int_0^{\infty} \frac{L_{\nu}(\nu_{\rm{e}})}{\nu_{\rm{e}}} R(\nu_e) d\nu_{\rm{e}},
\end{equation}
where $R(\nu_e)$ is the redshift-transformed filter response as a function of the emitted frequency and all other variables take their standard definitions. We leave any further cosmological and instrumental adjustments like aperture and surface brightness to the reader.

\begin{figure*}
\begin{center}
\includegraphics[scale=.26]{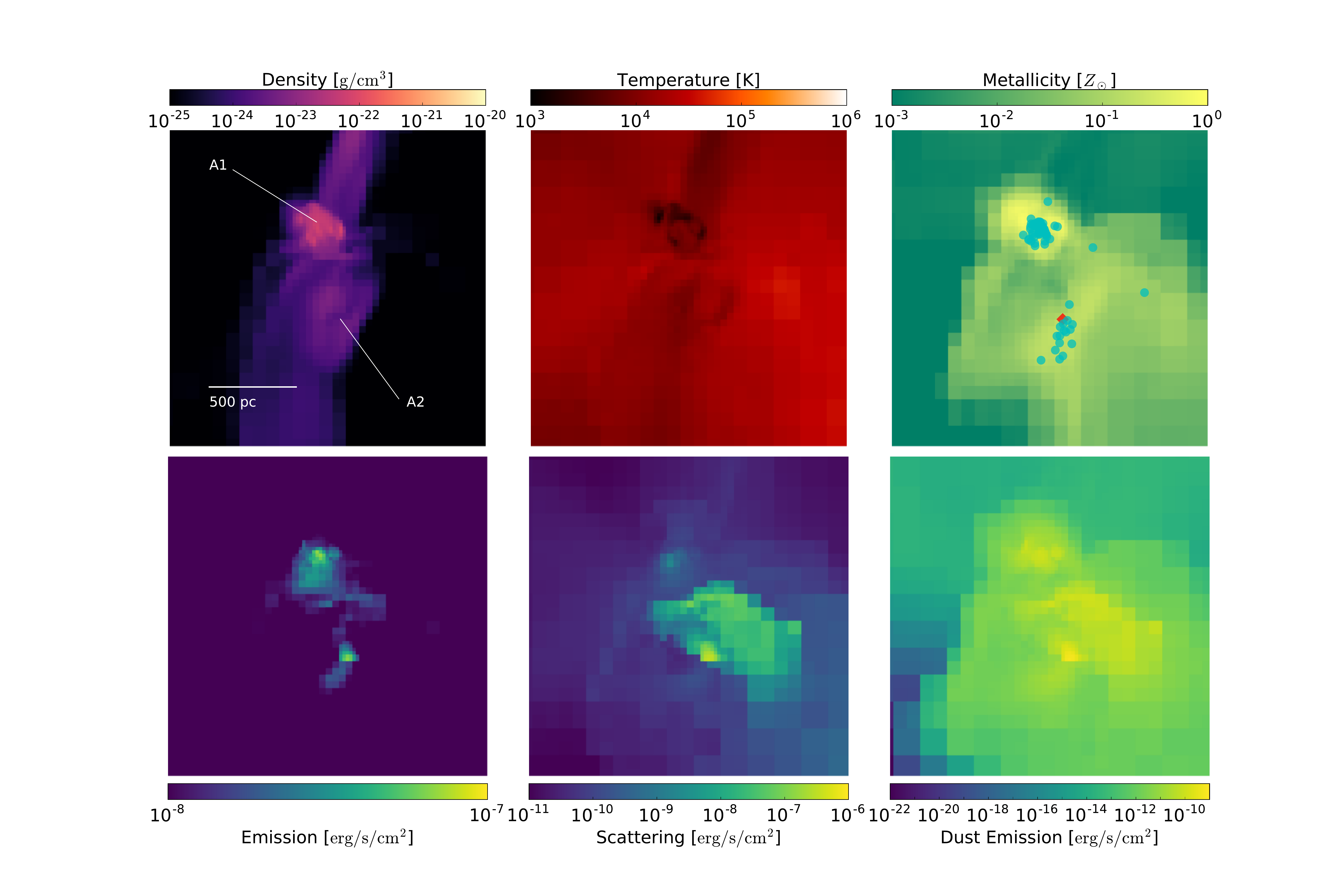}
\caption{Top row: Integral of density-weighted mean density (left), density-weighted mean temperature (middle), and density-weighted mean metallicity (right). Bottom row: Integrated total emission (left), scattering per unit area (middle), and integrated dust emission (right). The location of HMXBs are shown as diamonds. Circles are metal-enriched stellar clusters. Star markers are metal-free stars. Subhalos A1 and A2 are labelled in the top left plot for reference.}
\label{fig:scat}
\end{center}

\end{figure*}

\section{Results}
\label{sec:resul}

There are 146 halos in the ``rare peak" zoom-in region of the Renaissance Simulations with active metal-free stellar populations. As shown in Fig. \ref{fig:stellarplt}, most of these halos are small with a mean halo mass of only $3.40 \times 10^7\ {\rm M}_\odot$ owing to the tendency for these stars to form and die soon after a halo first cools into molecular clouds early in its evolution. Unfortunately, small halos imply small clusters of metal-free stars and low luminosity, therefore reducing the chance of a direct observation by any telescopes planned for the near future. However, sometimes mergers can mix stellar populations and generate scenarios where larger and brighter halos are influenced by ionizing photons from Population III stars and X-ray binaries, creating an opportunity to indirectly observe these objects sooner. More rarely, a Population III starburst may occur forming a relatively bright "Population III galaxy" \citep[e.g.][]{2009MNRAS.399...37J,2012AIPC.1480..101Z}.

Our simulation shows an example of both scenarios so we dedicate the first section of our discussion of our results on those two specific halos as well as the machinery of our radiative transfer pipeline. We then explore the emission lines trends and spectra of the full sample before finally presenting our photometric results.

\subsection{Stellar Population Merger Scenario}
\label{sec:merger}

\begin{figure*}
\begin{center}
\includegraphics[scale=.35]{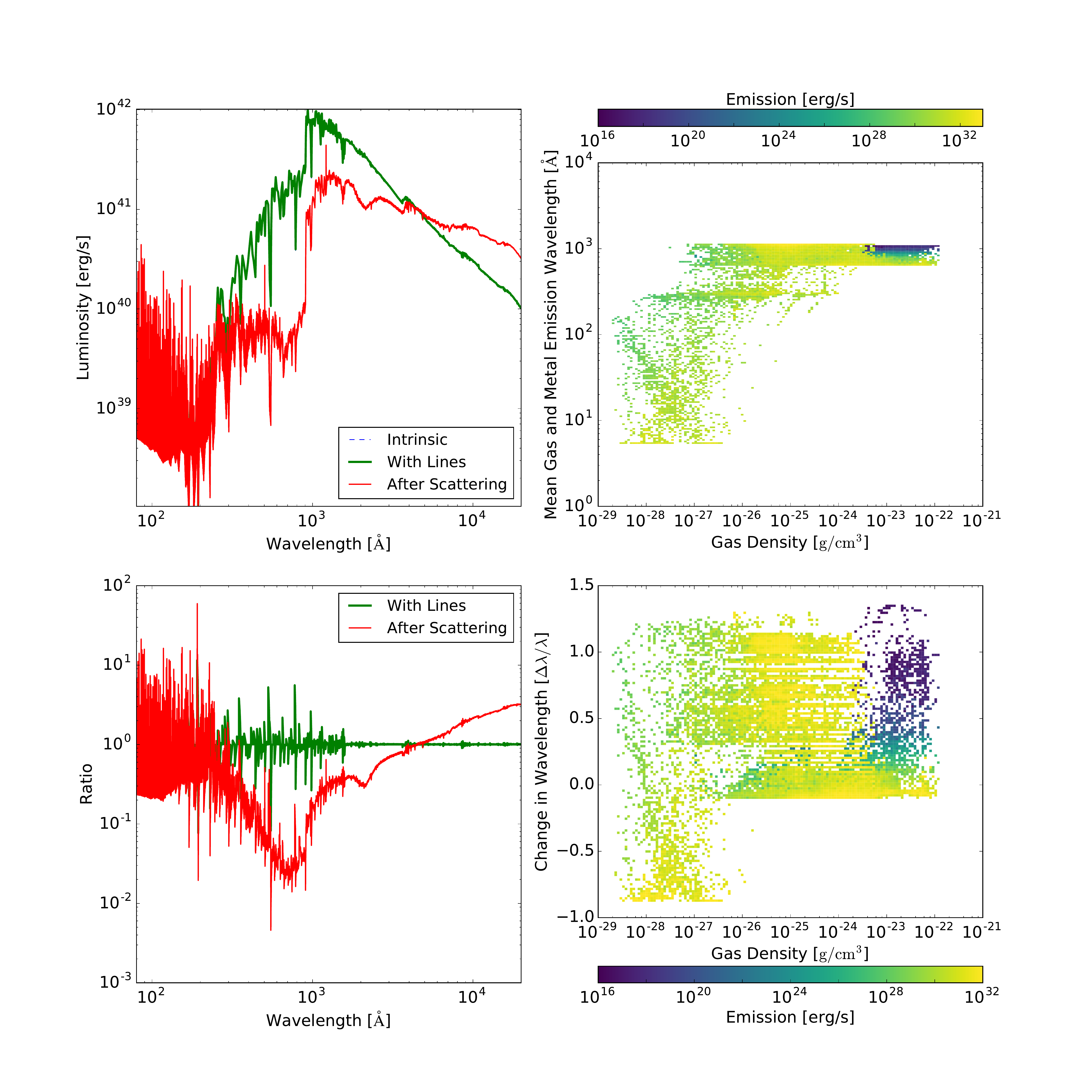}
\caption{Top row: The spectra of Halo A shown before the application of lines (dashed) with lines from gas in close proximity to the star (thick line) and after extinction (thin line). Here the dashed and think lines appear as nearly identical in the chosen scale. The plot on the top right shows the mean emission wavelength and power of the gas as a function of gas density. Bottom row:  Left-hand plot shows the ratio of emission to the nearest wavelength in the intrinsic spectrum. Some noise is present due to the relative coarseness of the intrinsic spectrum. The plot on the bottom right is the mean difference between the mean emission wavelength and the mean absorption wavelength for the combination of both gas and dust. }
\label{fig:phase}
\end{center}

\end{figure*}

\begin{figure*}
\begin{center}
\includegraphics[scale=.26]{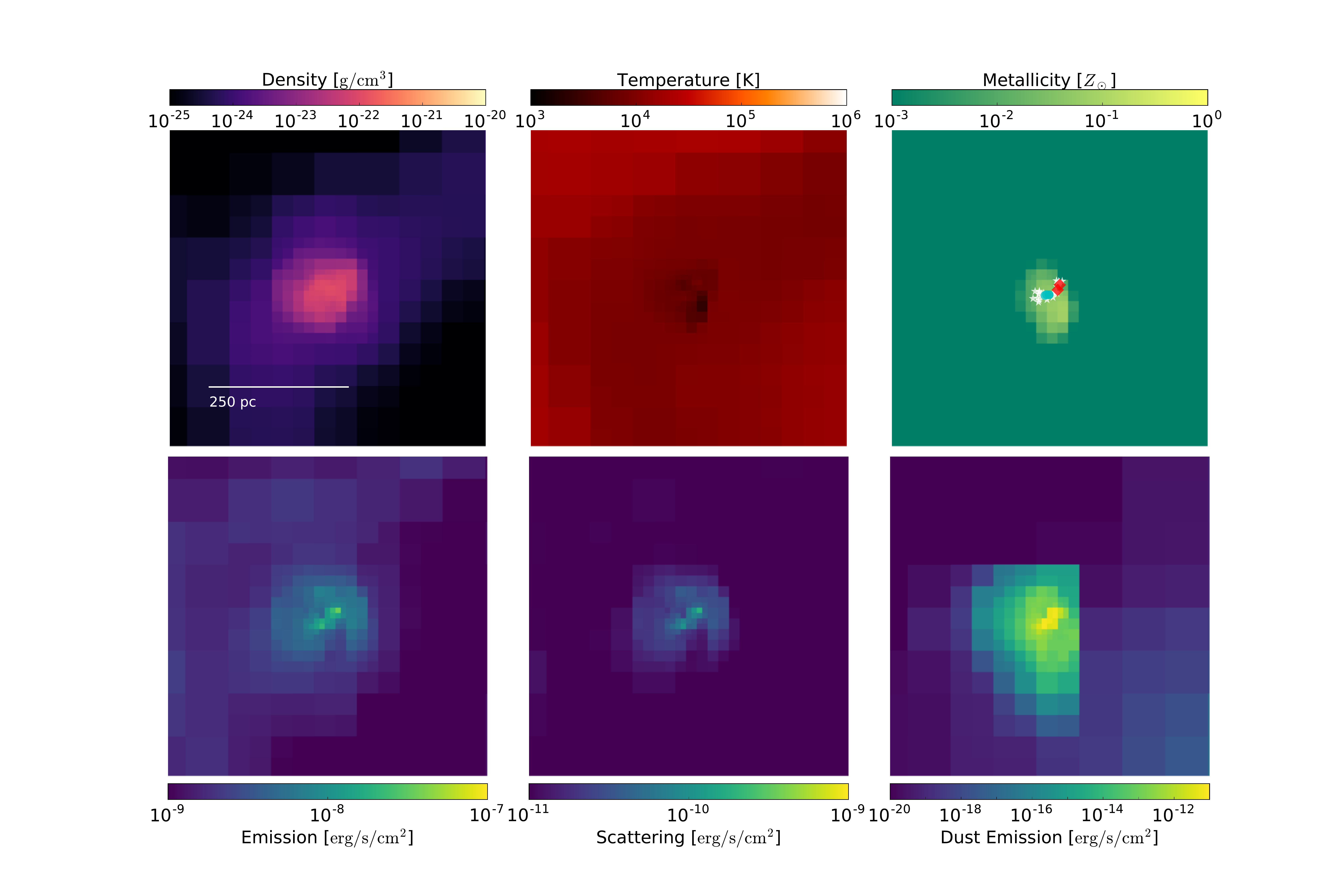}
\caption{A compact Population III stellar cluster plotted in the same manner as Fig. \ref{fig:scat}. The location of HMXBs are shown as red circles while individual Population III stars are shown in white. Cyan circles are metal-enriched stellar clusters.}
\label{fig:scat2}
\end{center}

\end{figure*}

\begin{figure*}
\begin{center}
\includegraphics[scale=.35]{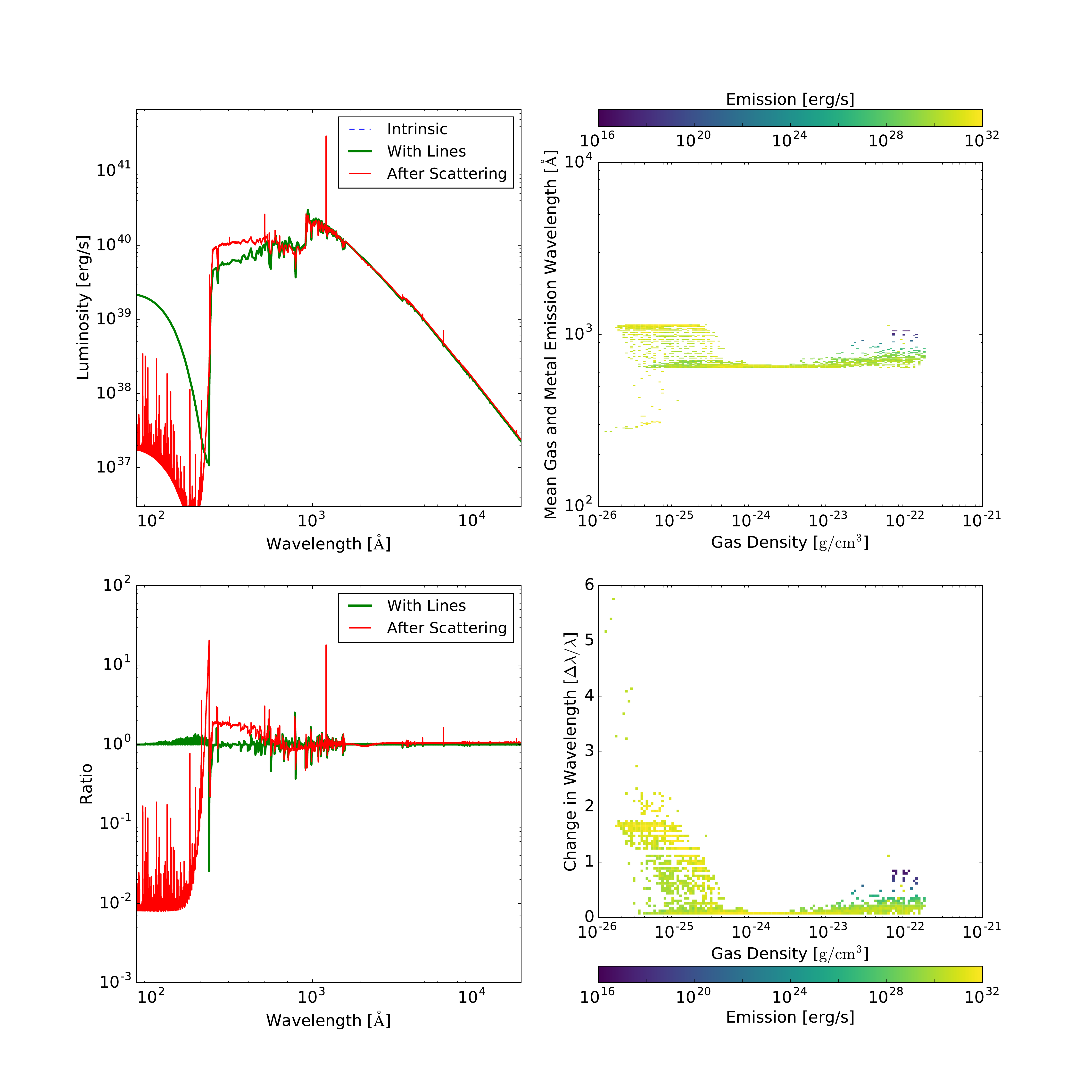}
\caption{Halo B plotted in the same manner as Fig. \ref{fig:phase}.}
\label{fig:phase2}
\end{center}

\end{figure*}

As described in Table \ref{tab:Halolist}, Halo A has a total mass of $1.30 \times 10^{8}\ {\rm M}_\odot$, a metal-enriched stellar population of $5.47 \times 10^{5}\ {\rm M}_\odot$, and two Population III stars totalling $6\ \rm{M}_\odot$ which we treat as high-mass x-ray binaries. The halo is composed of a compact, dense ($> 10^{-23}\ \rm{g\ cm^{-3}}$) clump (sub-halo A1) merging with a larger, lower density clump (sub-halo A2) as shown in the density weighted projections in the top row of Fig. \ref{fig:scat}. Sub-halo A1 hosts cool, metal-enriched gas while the sub-halo A2 is on average an order of magnitude hotter and hosts both metal-enriched and metal-free gas. The hottest gas ($T > 3 \times 10^4$ K) is concentrated in the CGM and in a supernova remnant to the right of the sub-halos as viewed in the figure. Sub-halo A2 contains the metal-free stars and by extension the HMXBs.

We integrate emission and scattering per unit volume from dust and gas along the projection axis used to produce the density, temperature, and metallicity figures resulting in plots of integrated specific luminosity in units of $\rm{erg\ s^{-1}cm^{-2}}$ which is proportional to but not equivalent to flux. Under the assumption of local thermodynamic equilibrium, the combination of dense, cool, and metal/dust-rich gas in sub-halo A1 results in thermal emission on the order of $10^{-8}$ to $10^{-7}\ \rm{erg\ s^{-1}cm^{-2}}$. The emission contribution from dust peaks at $\sim 4 \times 10^{-10}\ \rm{erg\ s^{-1}cm^{-2}}$ within a burst metal-enriched stars in close proximity to the HMXBs. Though this region has a lower metallicity than the peaks found in A1 as well a lower density of dust, warmer temperatures contribute to a higher overall dust emission.

To estimate scattering energy, we calculate the mean absorption-weighted albedo using the intrinsic stellar spectra and the density fraction of the constituents in the relationship

\begin{equation}
\langle \alpha \rangle_{x} = \frac{\int  \kappa_{\nu,x} I_\nu \alpha_{\rm{x}} d \nu}{\int \kappa_{\nu,x} I_\nu d \nu}
\end{equation}

\begin{dmath}
\langle \mathscr{E} \rangle_{\rm{scattering}} \approx f_{\rm{gas+metals}}\mathscr{E}_{\rm{gas+metals}} \frac{\langle \alpha \rangle_{\rm{gas+metals}}}{1-\langle \alpha \rangle_{\rm{gas+metals}}} + f_{\rm{dust}}\mathscr{E}_{\rm{dust}} \frac{\langle \alpha \rangle_{\rm{dust}}}{1-\langle \alpha \rangle_{\rm{dust}}},
\end{dmath}
where $I_\nu$ is the incident frequency-dependent flux. We find that the scattering in sub-halo A1 is limited to less than $10^{-9}\ \rm{erg\ s^{-1}cm^{-2}}$ while scattering in sub-halo A2 is of the order of $10^{-9}$ to $10^{-7}\ \rm{erg\ s^{-1}cm^{-2}}$. Taken together, this implies that the reprocessing of the intrinsic spectra in A1 is absorption and emission-dominated while the reprocessing of the high mass X-ray binaries in A2 is a roughly equally-weighted combination of gas emission, absorption, and scattering.

\begin{figure*}
\begin{center}
\includegraphics[scale=.38]{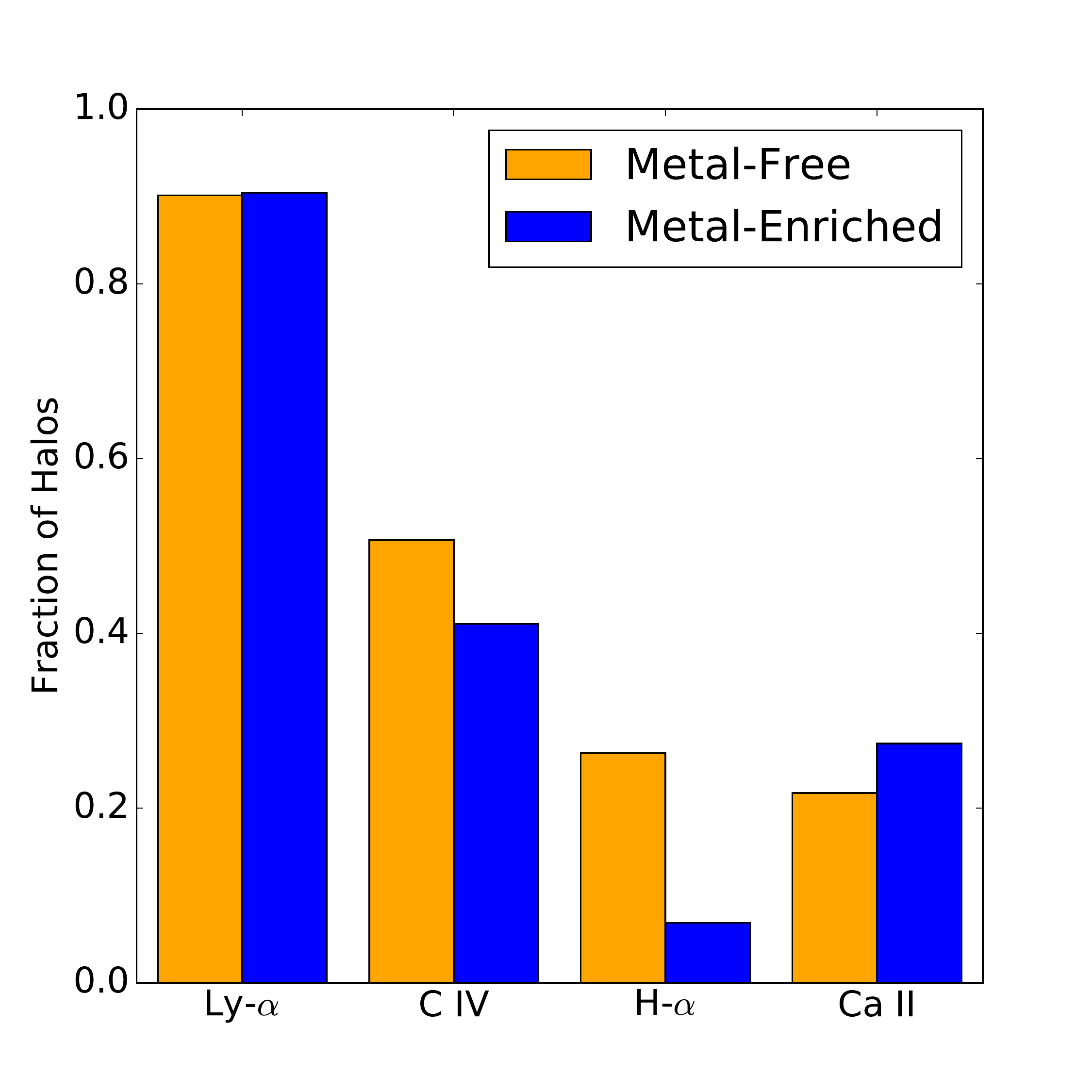}
\hfill
\includegraphics[scale=.38]{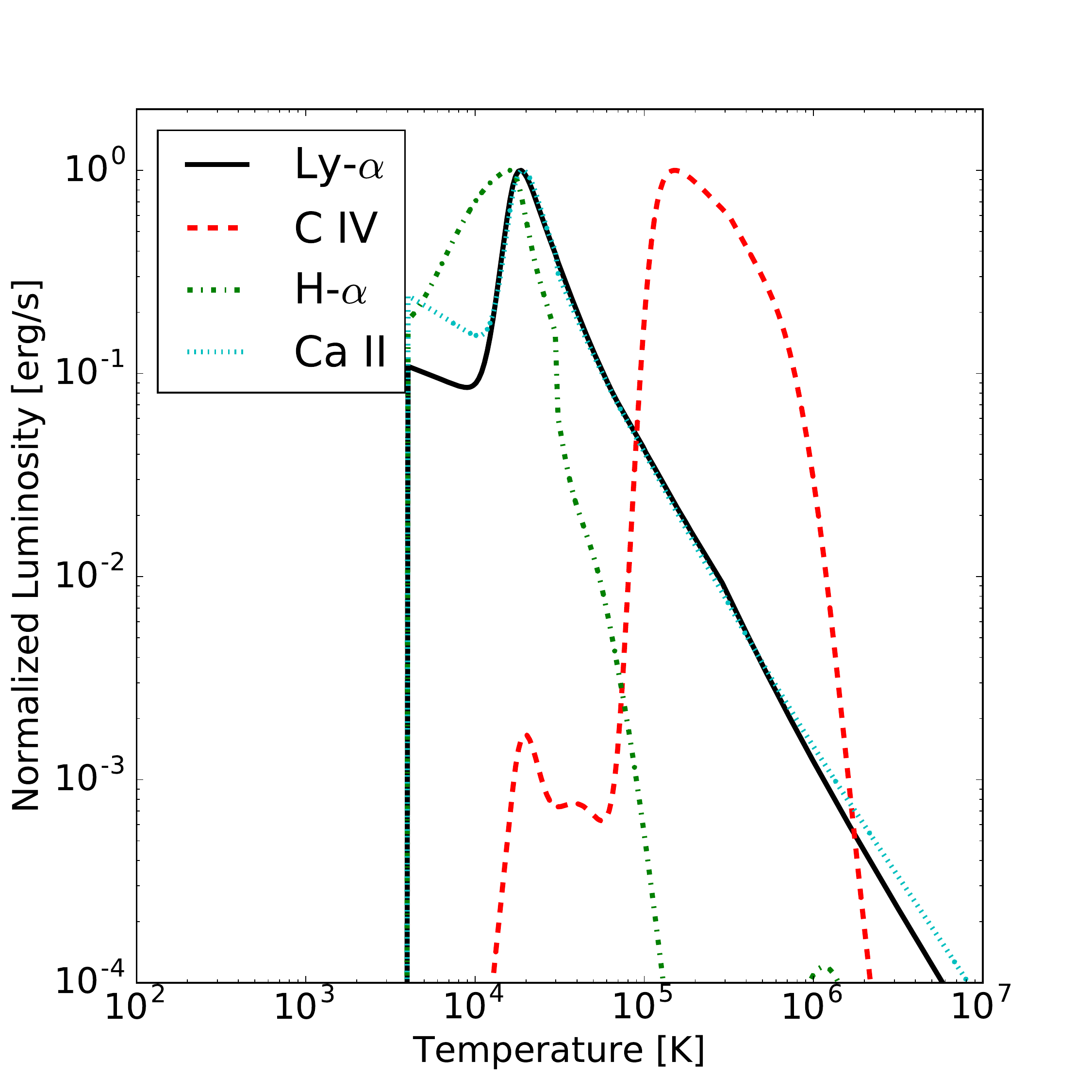}
\caption{Left: Four most common emission lines amongst halos in our simulation with active Population III stars (left bar) and only metal-enriched stars (right bar). Right: Emission versus  gas temperature normalized by the maximum emission of each line calculated for a fixed density and metallicity of $10^{-25}\ \rm{g\ cm^{-3}}$ and 0.1 $\rm{Z_\odot}$ respectively. C IV refers to the $\lambda\lambda1548,1551$ UV doublet and Ca II refers to the $\lambda\lambda\lambda8498,8542,8662$ IR triplet.}
\label{fig:lines}
\end{center}

\end{figure*}

As shown in Fig. \ref{fig:phase} (top left and bottom left) there are few differences between the intrinsic spectra and the spectra with local emission lines since the regions used to calculate those lines are compact. However, the impact of extinction of the spectra through the rest of the ISM is very pronounced. High-energy photons from the HMXB are reprocessed into emission lines by the metal-enriched gas while hydrogen-ionizing radiation is absorbed and reprocessed into IR photons. Lyman-Werner absorption is also pronounced except at Ly$\alpha$ where a high equivalent-width line forms from excitations in neutral hydrogen within sub-halo A1.  Thus, the spectra of Halo A demonstrates the signature of a cool metal-rich halo merging with a warm metal-free halo.

The components contributing to the reddening of the spectra is further demonstrated when we segregate the emission from gas and plot it against density and mean emission wavelength (Fig. \ref{fig:phase} top right). The hottest gas ($>10^{7}$ K) forms an artificial ridge at $\sim 5.3\ \AA$ at the limit of our calculation of the emission profile (Fig. \ref{fig:plotfour}) and a second ridge at $1.14 \times 10^{3}\ \AA$ at our minimum temperature of 316 K 
Between these two bounds we see that the lowest-density gas tends to be hotter and therefore more luminous per unit mass as we would expect. These bins correspond to the large number of emission lines seen at the high energy end of the post-extinction spectra. However most of the emission is coming from the cooler, medium density gas in A1. As mentioned, dust emission is weak, but present throughout the various density and temperature environments in the halo and therefore we see dust contributing to the spectra over the entire range of wavelengths simulated. The reddening plot (Fig. \ref{fig:phase} bottom right) shows the relative change in wavelength between absorption and emission. Like the scattering plot (Fig. \ref{fig:scat}, bottom center), absorption wavelength is an absorption weighted mean using the intrinsic spectrum. Therefore the mean change is

\begin{equation}
\langle \nu\rangle_{\rm{absorption},x} = \frac{\int  \kappa_{\nu,x} I_\nu \nu d \nu}{\int \kappa_{\nu,x} I_\nu d \nu},
\end{equation}

\begin{equation}
\langle \nu\rangle_{\rm{emission},x} = \frac{\int j_{\nu,x} \nu d \nu}{\int j_{\nu,x} d \nu},
\end{equation}
\begin{dmath}
\Delta \lambda  \approx f_{\rm{gas+metals}}\mathscr{E}_{\rm{gas+metals}} (\langle \lambda \rangle_{\rm{emission},\rm{gas+metals}}-  \langle \lambda \rangle_{\rm{absorption},\rm{gas+metals}}) + f_{\rm{dust}}\mathscr{E}_{\rm{dust}} (\langle \lambda \rangle_{\rm{emission},\rm{dust}}-  \langle \lambda \rangle_{\rm{absorption},\rm{dust}}).
\end{dmath}

The relative change, $\Delta \lambda/\langle \lambda \rangle_{\rm{absorption}}$, shows three distinct phenomena. At the low density end, we see that the hot gas has the propensity to either increase or decrease the wavelength of the spectra for mean absorption wavelengths $\sim 600\ \AA$. For moderate densities, all the gas contributes to the reddening of the spectra, and at high densities the contribution from high-density pockets of cool gas is seen to cause low-power reddening. Since the analysis of emission and absorption are drawn analytically from bulk characteristics, they do not paint a complete picture of reprocessing of the spectra. The impact of scattering and iterative energy balance in the Monte Carlo calculation as well as the precise three-dimensional distribution of gas, metals, and dust accounts for the difference between the intuition gained in the phase and projection plots and the fully simulated result in Fig. \ref{fig:phase} (top left). However, we predict the signature of the merger scenario to be markers of very high and relatively low energy gas and metals within the same halo.

\subsection{Population III Galaxy Scenario}
\label{sec:pop3}
\begin{figure*}
\begin{center}
\includegraphics[scale=.35]{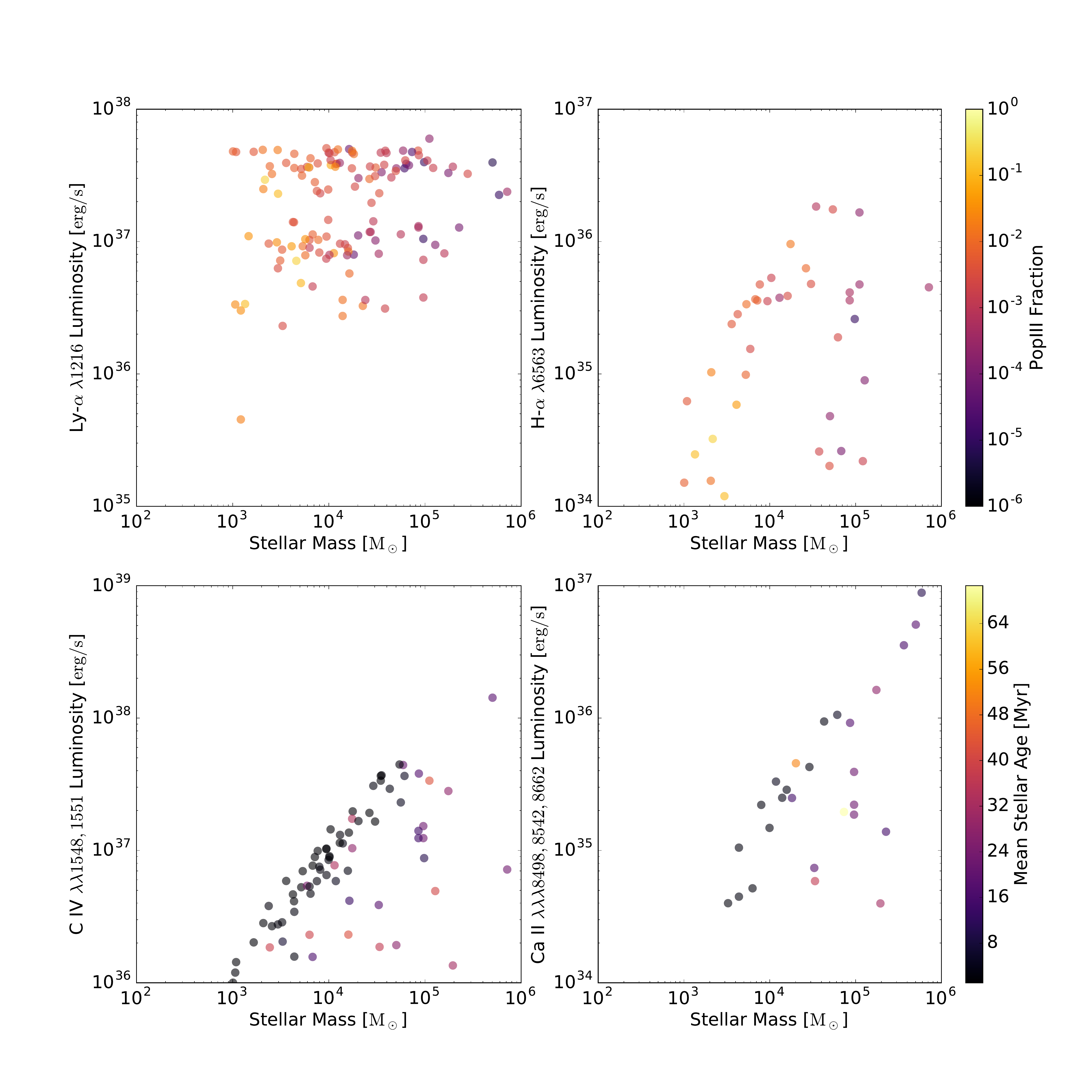}
\caption{Top row: Total luminosity of Ly$\alpha$ (left) and H-$\alpha$ (right) tinted by the ratio of the mass of Population III stars to the total stellar mass. Bottom row: C IV doublet (left) and Ca II triplet (right) luminosities tinted by mean stellar age, inclusive of both metal-free and metal-enriched stars. C IV is a product of an hot plasma and a high ionizing flux. Ca II lines have been examined in the broad line regions of AGN and in stellar atmospheres.}
\label{fig:lines2}
\end{center}

\end{figure*}

\begin{figure*}
\begin{center}
\includegraphics[scale=.35]{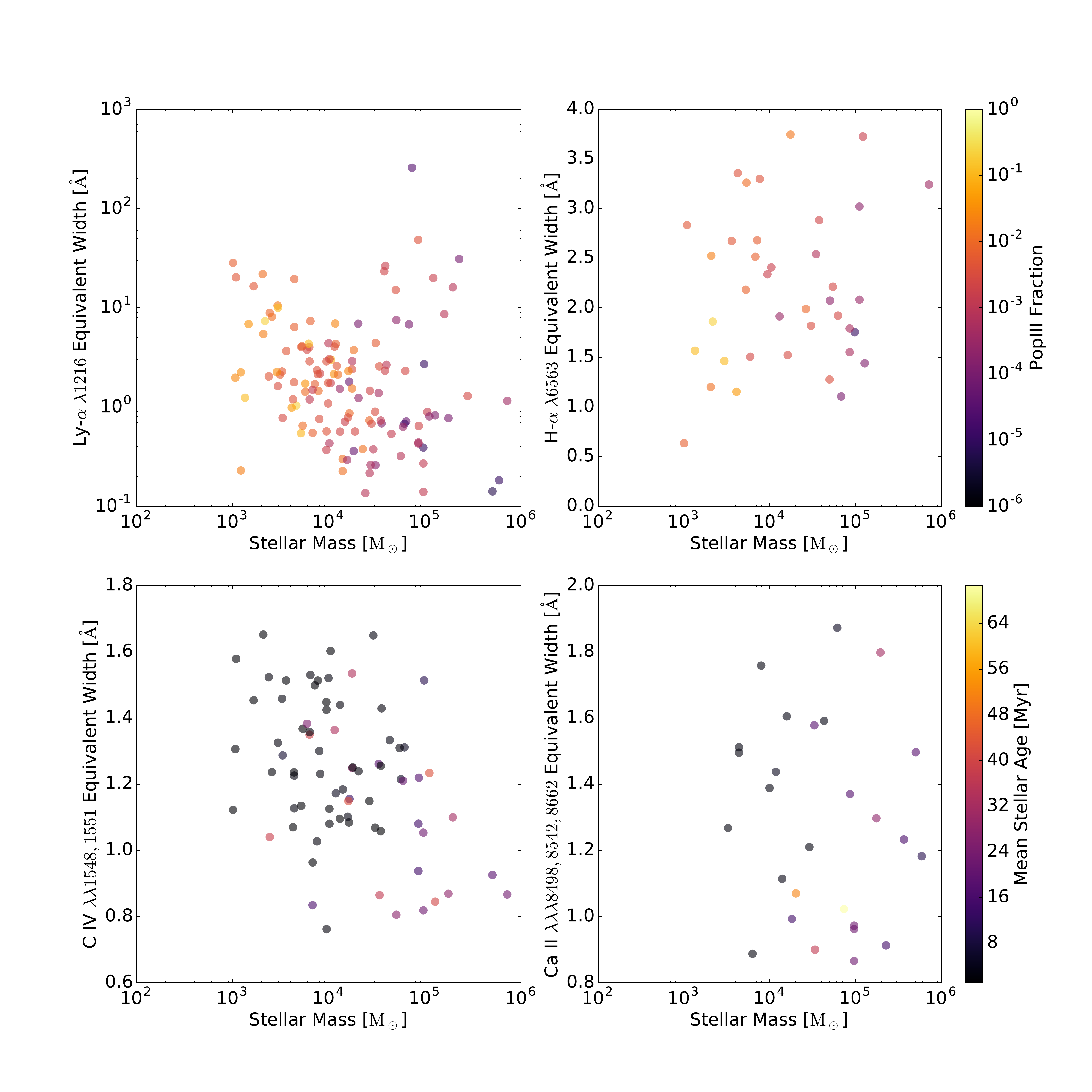}
\caption{Same as Fig. \ref{fig:lines2} but showing equivalent width rather than luminosity. Ly$\alpha$ equivalent width shows a negative log-log relationship with increasing stellar mass as a product of a nearly flat luminosity-mass relationship.}
\label{fig:lines3}
\end{center}

\end{figure*}

Halo B  has a total mass of $2.30 \times 10^{7}\ {\rm M}_\odot$, 9144 ${\rm M}_\odot$ of metal-enriched stars with an intrinsic bolometric luminosity of $4.38 \times 10^6\ {\rm L}_\odot$. There are an additional 30 Population III stars totalling 712 ${\rm M}_\odot$. Once again, we convert two of those stars into HMXBs with a 40 ${\rm M}_\odot$ compact object accreting at the Eddington rate. Combined, the HMXBs and Population III stars have an intrinsic bolometric luminosity of $4.79 \times 10^6\ {\rm L}_\odot$ of which the HMXBs contribute $2.72 \times 10^6\ {\rm L}_\odot$ mostly in soft and hard X-rays. Thus the intrinsic spectra is dominated by HMXBs in the X-ray, metal-free stars in the hydrogen-ionizing UV band, and metal enriched stars at higher wavelengths. 

As shown in Fig. \ref{fig:scat2}, allof the metal-free stars reside in the sole clump in a region of high density ($> 10^{-22} \rm{g\ cm^{-3}}$) gas, high metallicity, and relatively low temperature gas ($<10 ^4$ K). The range of temperatures are particularly remarkable because they imply that this cluster of Population III stars, the second largest in the entire simulation, has yet to undergo significant heating from ionizing radiation and a disruptive supernova. The cool gas results in absorption-dominated spectra reprocessing and low emittive power renders dust essentially irrelevant for our radiative transfer routines in this halo. The metal-enriched stars are embedded within the cluster of metal-free stars and have a mean age of only $3.5 \times 10^5$ yr and a mean metallicity of $0.17\ {\rm Z}_\odot$ after quickly being triggered by the death of a Population III star in the cool medium. 

The simple structure of Halo B results in two distinct signatures in the spectra as seen in Fig. \ref{fig:phase2} (top left and bottom left). While Halo A reprocessed emission from the HMXB into emission lines at roughly the same wavelength, we see that Halo B absorbs those photons and re-emits them as hydrogen-ionizing radiation. Also unlike Halo A, ionizing radiation is a strong component of both the intrinsic and reprocessed spectra and stronger than in the intrinsic spectra ($f_{\rm{esc}} >1$). Fig. \ref{fig:phase2} (top right) confirms that all of the emission is confined to the UV. The estimated mean change in wavelength due to absorption and re-emission (bottom right) is reddening by as much as factor of 7 $\Delta \lambda/\lambda$.

Of the emission lines, Ly$\alpha$ is again the most prominent UV emission line feature and about the same luminosity as the Ly$\alpha$ emission in Halo A despite having a lower bolometric luminosity by about an order of magnitude. This implies that Ly$\alpha$ alone cannot be used to distinguish between these two scenarios, but in the presence of H-$\alpha$ and the Ca II $\lambda\lambda\lambda8498,8542,8662$ triplet in the final spectra result from the cooler gas in Halo B.

\subsection{Emission Lines}

Since emission lines are generated as part of a Monte Carlo calculation, no $a\ priori$ knowledge of the continuum emission can be used to identify the lines or calculate their equivalent widths. Therefore we subject the SEDs to a series of tests to determine the presence, source, and peak wavelengths of the lines based on the first and second derivatives of the continuum. This includes a check that eliminates line candidates with equivalent widths below 0.75 \AA\ unless the peak is two or more times the continuum, which has the effect of removing weak, but present features from our study so our counts should be interpreted as lower bounds. We also restrict our line catalogues to rest wavelengths greater than the Lyman limit.

The most common emission lines in order of their frequency within our sample are the Ly$\alpha$ $\lambda$ 1216 line, the C IV $\lambda\lambda1548,1551$ doublet lines, the Balmer $\alpha$ (H-$\alpha$) line, and the Ca II $\lambda\lambda\lambda8498,8542,8662$ triplet lines which appear in the spectra of 90\%, 51\%, 26\%, and 22\% of the 146 halos respectively. To construct a control group, we found the closest match between the intrinsic JWST NIRCam F277 wideband filter of a half-sample of halos with metal-free stars to the intrinsic flux of 73 halos without metal-free stars or HMXBs within the ``rare-peak" simulation at the same redshift. Within that group, Ly$\alpha$ was detectable in 90\%, the C IV $\lambda\lambda1548,1551$ doublet in 41\%, H-$\alpha$ in 7\%, and Ca II $\lambda\lambda\lambda8498,8542,8662$ in 27\%, as shown in Fig. \ref{fig:lines} (left). This implies that H-$\alpha$ is significantly more rare or weak in the spectra of our control group than in our sample of halos with metal-free stars and HMXBs.

\subsubsection{Lyman-$\alpha$}

Substantial Ly$\alpha$ emission is prominent in gas with temperatures between $10^4$ and $10^5$ K. Because these temperatures are associated with warm H~{\sc II} regions and Population III stars that are generally short-lived and embedded in their birth clouds, higher Ly$\alpha$ emission is a feature in most of our SEDs. Unlike the other common emission lines, we see fairly consistent Ly$\alpha$ luminosity between $10^{36}$ and $10^{38}$ $\rm{erg\ s^{-1}}$ in halos with high and low fractions of Population III stars as well as high and low masses of metal-enriched stars (Fig. \ref{fig:lines2}, top left). Consequently, there is a mostly well-correlated inverse relationship between equivalent width and both total stellar mass as well as Population III fraction (Fig. \ref{fig:lines3}, top left). As shown in \cite{2017MNRAS.469.4863B}, the mass-weighted mean age of metal-enriched stars in halos with total stellar masses between $10^5$ and $10^6$ ${\rm M}_\odot$ begin to settle into a narrow range as the H~{\sc II} regions around stars no longer encompass the entire halo. This allows star formation to transition from a series of bursts to a pattern of continuous formation. While emission continues from star-forming regions, the average stellar cluster is older and divorced from its birth molecular cloud and thus Ly$\alpha$ emission does not scale with stellar mass. Secondarily, the extinction cross section of Ly$\alpha$ photons in neutral hydrogen is high, implying that scattering through dense gas may attenuate emission along any line of sight.

While our prior treatment shows that the inverse relationship between intrinsic Ly$\alpha$ equivalent width (EW) and total stellar mass is generally extensible to metal-enriched stellar populations in similar environments, below $10^4$ ${\rm M}_\odot$, the pattern is exclusive to starbursts as several halos have had their star formation extinguished and have low intrinsic Ly$\alpha$ EW. For our sample, high metal-free stellar mass to total stellar mass fractions consistently exhibit the highest Ly$\alpha$ EWs (> 5 \AA) in our sample after extinction.

\subsubsection{$C\ IV\ \lambda\lambda1548,1551$}

C IV UV emission lines are an intrinsic feature of active galactic nuclei \citep[AGN; see review by][]{2000A&ARv..10...81V}. Their inclusion in our results demonstrate the consideration of line transfer from high-energy interactions to lower-energy photons in our calculations. Due to the ionization energies of carbon, C IV $\lambda\lambda1548,1551$ emission requires temperatures between $10^5$ and $10^6$ K. In this temperature range, absorption cross sections are highest for X-rays in metal-enriched gas, so the presence of high-energy sources like HMXBs directly impacts the prominence of this line. The occurrence of the C IV UV emission lines in approximately half of the halos in our sample is therefore partially a product of our decision to affix a HMXB to almost all of our halos. 

In environments with low-metallicity ISMs and CGMs like those we see in our sample, X-ray escape fractions are generally high so despite a fixed number of HMXBs, the doublet's luminosity grows proportionally to the mass of the halo as seen in Fig. \ref{fig:lines2} (bottom left). The doublet is also mostly confined to halos with very young stellar populations (< 24 Myr) as these halos are more likely to have hot gas heated either by supernovae or photo-heating from young stars. This is in contrast with Ly$\alpha$ emission which also implied young stellar populations but required the persistence of colder star-forming gas.

C IV UV doublet equivalent widths were consistently between about 0.8 and 1.7 \AA\ as their strength is closely tied to absorption of the incident spectra which, in conjunction with the availability of metal-enriched gas, caps their equivalent widths.

\subsubsection{H-$\alpha$}

The Balmer series in ionized hydrogen forms from a recombination cascade in diffuse medium. This is tempered by collisional excitations in warm gas where the density and energy of particles or photons are high enough to ensure continuous re-ionization. Balmer-series emission is more susceptible to this effect than Ly$\alpha$ due to the lower energy of their transitions. Thus the relative luminosity of these lines both peak and drop off at lower temperatures.

H-$\alpha$ luminosity scales with stellar mass and the fraction of Population III stars in our sample.  As shown in Fig. \ref{fig:lines}, H-$\alpha$ emission implies gas temperatures below $5 \times 10^4$ K and by extension the coolest star-forming halos and molecular clouds. However unlike Ly$\alpha$ emission, H-$\alpha$ emission has a higher escape fraction in neutral hydrogen and is therefore less susceptible to attenuation in dense gas. As shown in Fig. \ref{fig:lines3} (top-right), H-$\alpha$ EW also scales weakly with stellar mass and is the only one of the prominent emission lines to do so in agreement with observations of H$\alpha$-derived specific star formation rates of higher mass galaxies at $z\sim 2$ by \citet{2006ApJ...647..128E}. Higher fractions of metal-free stars are roughly inversely related to  H-$\alpha$ EW which is a function of both the overall tendency for Population III stars to be a smaller fraction of the stellar mass in larger halos and heating from metal-free stars. With $JWST$, H-$\alpha$ emission for objects at $z=15$ should appear in the MIRI F1000W band. We note that IR Paschen-$\alpha$ transitions at 18750 \AA\ were also present in 33 halos due to the recombination cascade.

\subsubsection{$Ca\ II\ \lambda\lambda\lambda8498,8542,8662$}

The ionization potential of Ca$^+$ is $\sim$ 11.9 eV, making  it susceptible to ionization by strong Ly$\alpha$ (10.19 eV) due to the presence of a meta-stable energy state of 1.7 eV Ca$^+$ that provides the difference \citep{1989A&A...208...47J}. Thus the Ca II NIR triplet emission neatly overlaps the thermal trends of Ly$\alpha$ and is only weakly related to temperature in its absence. Therefore, like Ly$\alpha$, Ca II NIR triplet emission is tied to AGN \citep{1989ApJ...347..656F} and bursts of star-formation in metal-enriched gas \citep{1993Ap&SS.205...85G}. However, as shown in Fig. \ref{fig:phase2}, gas metal-enrichment within large Population III stellar clusters is sufficient to generate this line so its presence does not automatically indicate a metal-enriched stellar population. We observe a well-correlated power law relationship between the emission of this triplet and the luminosity of the halo and no discernible relationship in equivalent widths. These trends imply that halos are mostly transparent to Ca II emission and the precise arrangement of the gas, dust, and metals are less important than the incident flux and the temperature of metal-enriched gas. With $JWST$, Ca II emission for objects at $z=15$ should appear in the MIRI F1280W and F1500W bands. 

\begin{figure*}
\begin{center}

\includegraphics[scale=.35]{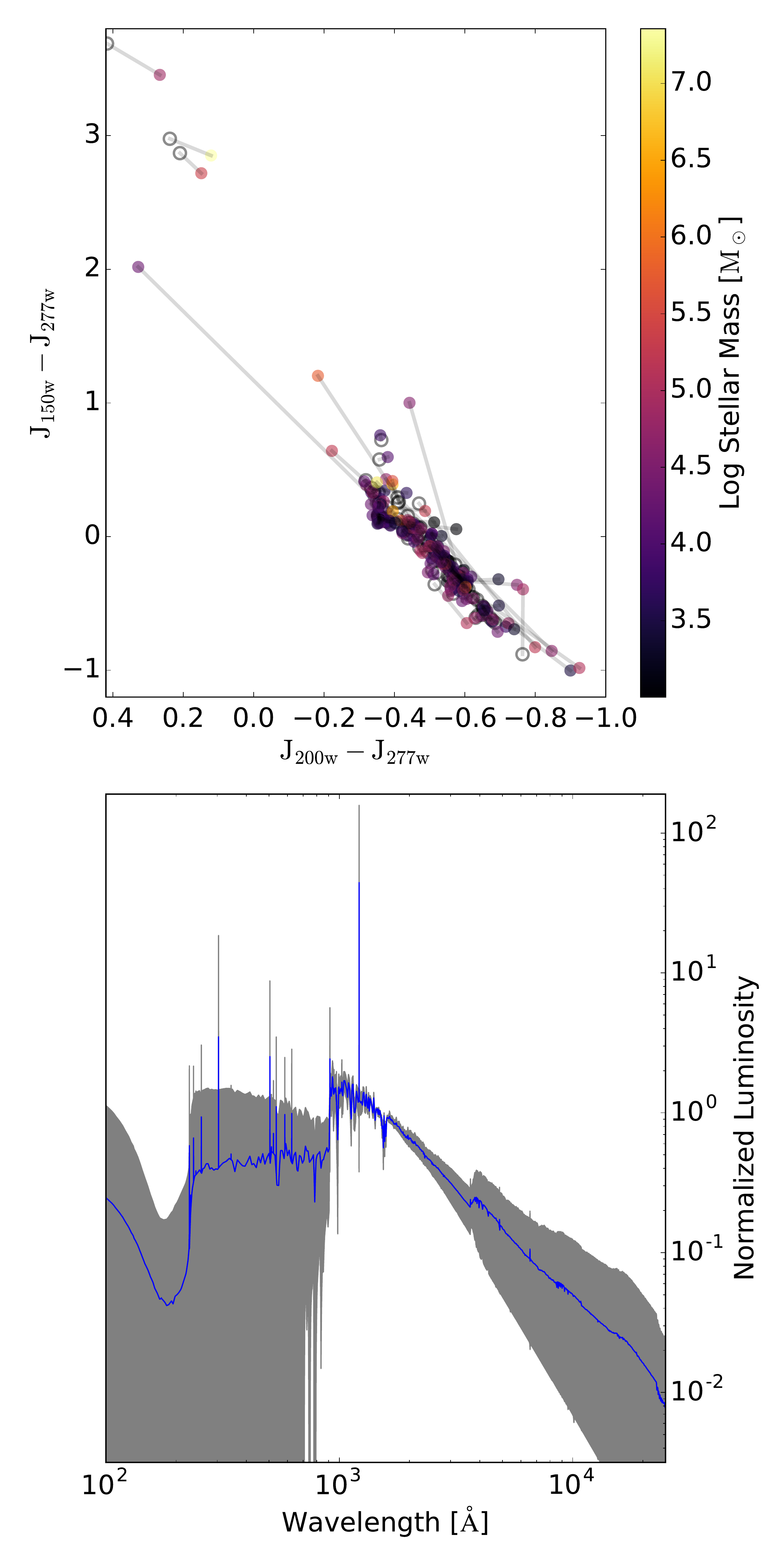}
\includegraphics[scale=.35]{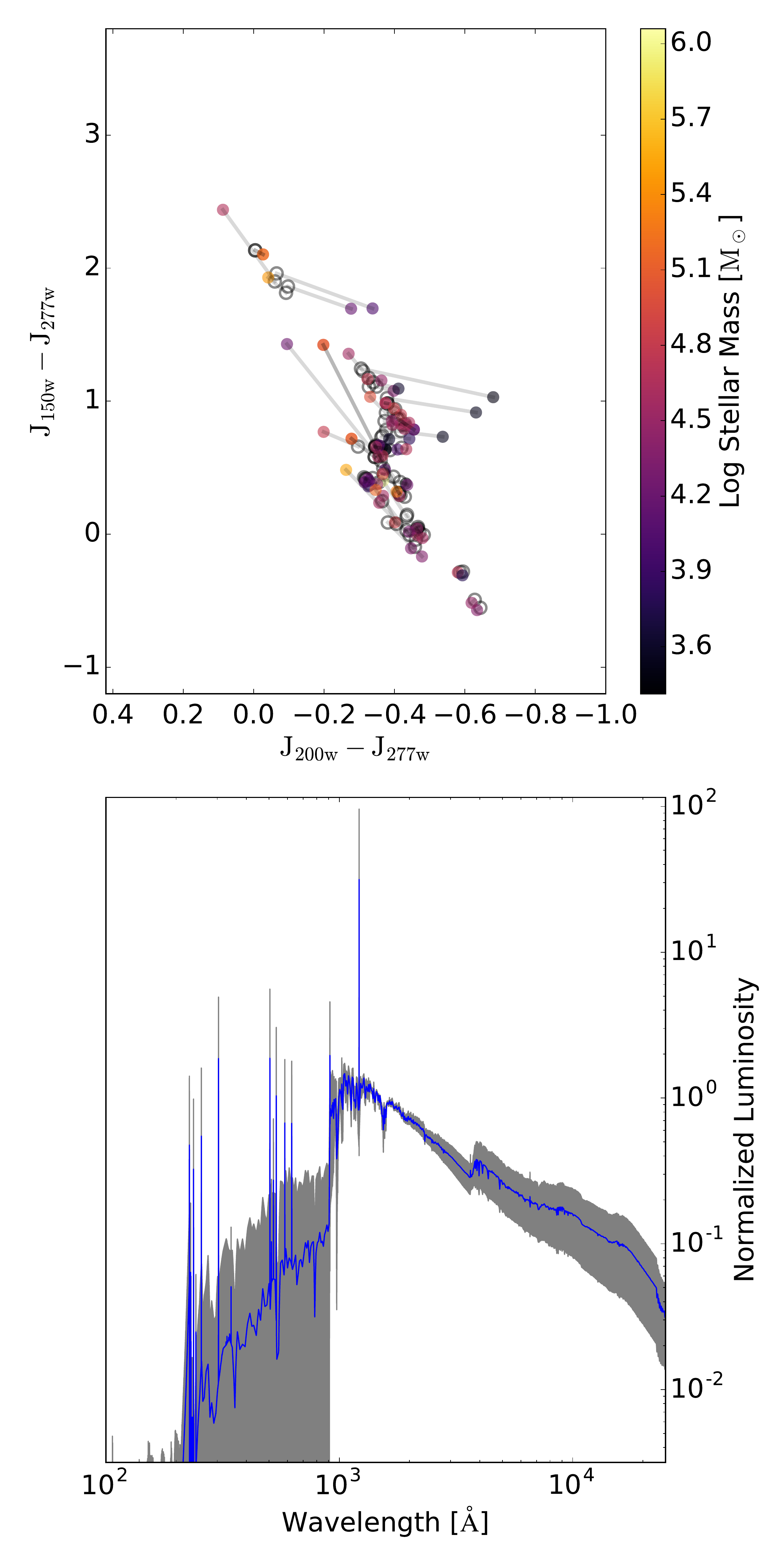}

\caption{The right-hand panels show a control sample of galaxies without metal-free stars or HMXB, but with the similar intrinsic $J_{277w}$ fluxes to our sample of galaxies with both (left-hand). Top: $JWST$ colour-colour plot of the sample of halos with at least  1 $\rm{M_\odot}$ Population III stars tinted by their total stellar masses at a redshift of  $z = 15$. Open circles and lines show changes from the intrinsic stellar and HMXB spectra. Bottom: Mean final spectral energy distribution ($\nu f_\nu$ vs \AA) of the sample shaded by one standard deviation above and below the mean and normalized by the values at 1500\AA.}
\label{fig:color-clolor}
\end{center}

\end{figure*}

\begin{figure*}
\centering
\includegraphics[scale=.16]{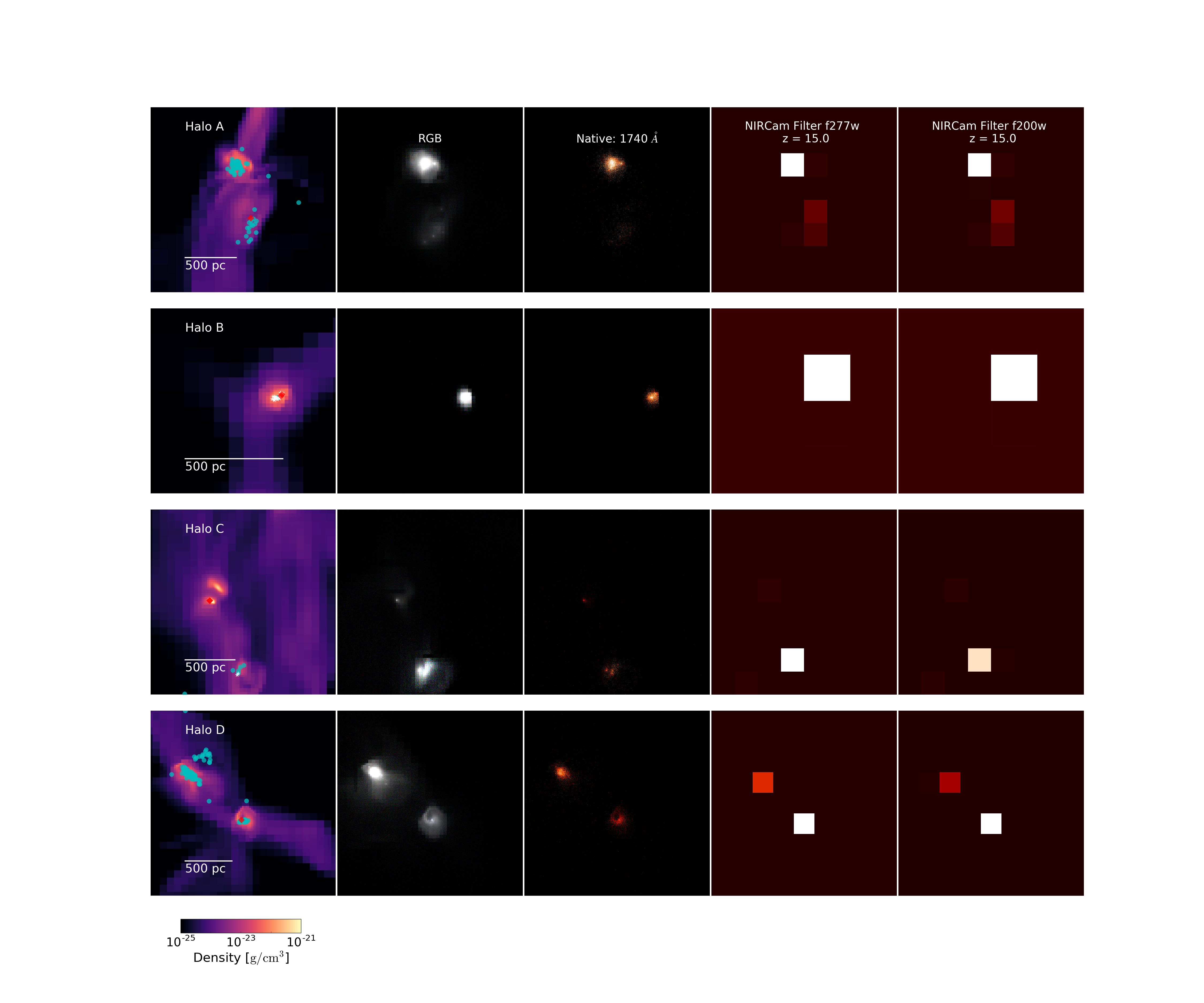}
\caption{Left to right: Mean density, rest frame optical RGB composite, monochromatic image corresponding to the $\rm{J_{277w}}$ image, $\rm{J_{277w}}$ image, and $\rm{J_{200w}}$ image for halo A, B, C, and D with a magnification factor of 10 and a 1 Ms exposure time. Markers in the density plots are circles, stars, and diamonds for metal-enriched stellar clusters, metal-free stars, and HMXBs respectively.}
\label{fig:photos}

\end{figure*}

\subsection{Aggregate Spectrographic and Photometric Results}

We produce $JWST$ colours by applying a filter throughput to our SEDs after accounting for the effects of redshifting. For our sample, we find that the intrinsic SED of our stars and HMXBs are poor predictors of the final colours produced by our radiative transfer calculations. As shown in Fig. \ref{fig:color-clolor} (top left), halos with final $\rm{J_{200w} - J_{277w}}$ and $\rm{J_{150w} - J_{277w}}$ colours around 0.45 and 0.25 respectively are likely to have changed little or reddened slightly after our calculations. Conversely, many halos with intrinsic colours in this range have their colours change drastically after processing. Shifts to a bluer colour implies reddening from higher energy photons from metal-free stars into the UV range of the filter and were mostly absent from the analysis of metal-enriched populations in the antecedent work. Drastic changes in $\rm{J_{150w} - J_{277w}}$ occur because the $\rm{J_{150w}}$ filter straddles the Lyman limit at $z =15$ so the colour is extremely sensitive to the production and escape fraction of ionizing radiation. The prominent Ly$\alpha$ lines at $\sim$ 19,450 \AA(observer) are captured by $JWST$ NIRCam's $\rm{J_{200w}}$ filter with the caveat that they are likely to be subject to extinction in the IGM not captured by our analysis. We expect this extinction to be particularly prominent early during the Epoch of Reionization and will seek to capture this effect in future studies. Here, the strength of this line increased $\rm{J_{200w} - J_{277w}}$ colours in many low stellar mass halos, sometimes dramatically.

\begin{table*}
  \centering
  \caption{Individual halo properties.}
  \begin{tabular*}{0.99\textwidth}{@{\extracolsep{\fill}} cccccccc}
    Halo & $\log M_{\rm tot}$ & $\log M_{\rm \star,ME}$ & $\log M_{\rm \star,MF}$ & $\log \rm{J_{277w}}$ & S/N$_{\rm max}$ & $\log \rm{J_{200w}}$ & S/N$_{\rm max}$ \\
    & [\Ms] & [\Ms] & [\Ms] & [$\rm{erg\ s^{-1}cm^{-2}}$]  & & [$\rm{erg\ s^{-1}cm^{-2}}$] & \\
    \hline
    A & $8.47$ & $5.74$ & $0.78$   &$-20.22$ & 3.92 & $-20.16$ & 5.08 \\
    B & $7.69$ & $3.96$ & $2.85$   &$-21.46$ & 1.03 & $-21.22$ & 1.02 \\
    C & $8.65$ & $5.49$ & $2.42$   &$-21.34$ & 2.71 & $-21.09$ & 2.91\\
    D & $8.56$ & $5.63$ & $0.48$   &$-20.04$ & 7.59 & $-19.87$ & 9.25\\
    \hline
  \end{tabular*}
  \parbox[t]{0.99\textwidth}{\textit{Notes:}
    The columns show halo mass, metal-enriched stellar mass, metal-free stellar mass, and $JWST$ $\rm{J_{200w}}$ and $\rm{J_{277w}}$ filter fluxes at $z=15$. Signal to noise ratios are for the brightest pixels shown in Fig. \ref{fig:photos} with a 1 Ms exposure time and $\mu = 10$.}
\label{tab:Halolist}
\end{table*}

Fig. \ref{fig:color-clolor} (bottom left) shows a composite of all the final SEDs in our sample normalized to emission at 1500 \AA(rest). In addition to the oft-mentioned Ly$\alpha$ line, we see several lines in the Lyman continuum. We did not explore these features because we expect neutral hydrogen to drastically attenuate ionizing emission in the IGM, but we note that several strong He I and He~{\sc II} emission lines are present in the ISM and CGM of most of the halos in our sample. In general, the fraction of ionizing radiation varied drastically between our halos due to the wide range of Population III stellar mass fractions. At the extremes, ionizing radiation was completely attenuated in some cases and represented the peak emitted energy in others. In the UV, results were more consistent as halos fell into the narrow range of 0.3 to 0.8 $\rm{J_{200w} - J_{277w}}$ demonstrated in the colour-colour plot with a few a outliers. 

The control group of 73 halos with similar intrinsic $J_{277w}$ flux, but composed of metal-enriched stars and no HMXBs are plotted in the right-hand panels of Fig. \ref{fig:color-clolor}. Colours show fewer outliers are generally more red in both colors. The aggregate spectra show more prominent emission lines from metals in the Lyman continuum and a shallower UV, visual, and infrared slope.

We explore photometry for the halos examined in Section \ref{sec:merger} and \ref{sec:pop3} as well as two more merger scenarios (Halos C and D). The composition and $JWST$ fluxes for all four halos are shown in Table \ref{tab:Halolist}. For our analysis, we take background noise to be Gaussian with a mean and standard deviation given by 

\begin{equation}
\langle N \rangle  = \frac{5 \times 10^{-8} \rm{Jy}}{\sqrt{t_{\rm exposure}}},
\end{equation}
which approximates the sensitivity of $JWST$'s NIRCam. We choose the lower band of the pixel colormap to be the maximum of our mean noise value and one standard deviation below the mean pixel flux which produces tinting of the low flux pixels in our processed images. We take the signal to noise ratio to be the brightest pixel divided by 2$\langle N \rangle$. 

Each halo is brighter than the noise through $JWST$ at $z = 15$ assuming a 1 Ms exposure time and a factor of 10 magnification using gravitational lensing. Generally however, galaxies of this kind are not resolved by $JWST$ and extend over only 1-2 pixels. The stellar population merger scenario (Halo A) occupies four pixels and appears as two distinct sub-halos of two pixels each. The Population III galaxy scenario (Halo B) illuminates a single pixel and produces a stronger flux in the $\rm{J_{200w}}$ filter than the $\rm{J_{277w}}$ filter, but we do not expect it to be directly observable with $JWST$ at this redshift without an extensive exposure time. Halo C however also features a high fraction of Population III stars and should be barely observable with a S/N of 3 in our example. Halo D would be the brightest of the four, but only contains a single Population III star amidst a much more luminous metal-enriched stellar population. 

\section{Discussion}
\label{sec:diss}

To facilitate our study of the spectrographic impact of Population III stars and HMXBs, we have contributed a few methodological improvements to radiative transfer post-processing of cosmological simulations. At the core of our calculations is the dust radiative transfer code {\sc Hyperion} which we extend to gas and emission lines by creating two dimensional arrays of extinction and emission prescriptions with {\sc Cloudy}. With irony, we note that our relatively simple treatment of dust leaves the most to be desired and improvements to our dust models will be part of the focus of future investigations. However, galaxy dust ratios at high redshift have been shown to vary greatly with the assumed grain accretion timescales  \citep{2015MNRAS.451L..70M} which itself varies sensitively with ISM density \citep{2016MNRAS.457.1842S} and composition, making dust difficult to constrain. 

We also briefly explore the prevalence of high mass X-ray binaries in Section \ref{sec:HMXB}. Since the impetus for those calculations was a desire to physically motivate their inclusion in our study, we were less concerned with the implications of their global fraction on the cosmological environment, but that subject certainly deserves some consideration. The multi-color accretion disk model implies an inverse relationship between black hole mass and peak temperature. This suggests that larger black holes emit more of their radiation as hydrogen-ionizing photons than smaller ones, which emit most of their energy at wavelengths too small to interact with gas in the ISM and CGM, but contribute to slow heating of the IGM for photons in the 500 eV to 1 keV energy range \citep{2014ApJ...791..110X}. This may have considerable implications for reionization, star formation, and estimates for escape fractions if luminous high mass compact objects are determined to be fairly prevalent. Furthermore, X-ray emission from binaries have been shown to strongly contribute to the cosmic X-ray background \citep{2016ApJ...833...84X}.

Generally, metal-free stars at high-redshift remain elusive to direct detection with their supernovae as the best chance of detection \citep[e.g.][]{2013ApJ...762L...6W}. Galaxies where Population III starbursts comprise most or all the stellar mass like Halo B were too small and dim to be observed with $JWST$ even with generous exposure times and gravitational lensing. In scenarios with a merger between a halo with a metal-enriched stellar population and a metal-free stellar population, the metal enriched population provides enough of a boost to the luminosity to make the halo discernible, but dense gas in deeper potential wells limits the permeability of ionizing and UV radiation. In this case, it may be possible to estimate the temperature profile from the UV slope and deduce the presence of a hot, ionizing source like a HMXB or a metal-free stellar cluster. However, the best and rarest scenario for observation was a merger between two galaxies with metal-free stars (Halo C), but there was only one such configuration in our simulation comoving box size of $133.6\  \rm{Mpc}^3$. Therefore, we predict that direct observation is possible at this redshift, but fairly improbable with the current generation of hardware.

In their analysis of the void region of the Renaissance Simulations, \citet{2016ApJ...823..140X} discover Population III star formation in the terminating redshift of $z = 7.6$ in halos that were generally larger than those hosting these stars in the rare peak volume. Late formation is enabled by strong LW flux from metal-enriched stars suppressing formation in the surrounding pristine gas and may continue to even lower redshift. Their sample includes rather large Population III starbursts with one in excess of $10^3\ {\rm M}_\odot$. There are twelve halos with active Population III stellar populations at the terminating redshift of the void simulation in a comoving volume of 220.5 $\rm{Mpc}^3$. Given their luminosity and redshift, some of these would likely be detectable with $JWST$.

Our use of averaged metal-free IMF prescriptions likely has little impact on observables like colour or imaging. By maintaining the size of the ionized region and temperatures from our simulation, the effect of this discrepancy is mostly minimized to calculation of the absorbed radiative energy within the Monte Carlo step. However since observation requires either a large number of metal-free stars or a metal-enriched population, the impact of small changes in the incident spectra of individual stars is vanishingly small, especially when compared to contributions from the other factors like the impact of morphology and viewing angle when observing irregular galaxies. For a detailed study of astrophysical radiative transfer phenomena on the other hand, a more robust spectral routine would be desirable. 

For objects at high redshift, emission line diagnostics serve as more of a long term prediction and a theoretical exploration with the notable exception of the Ly$\alpha$ line, which sits near the center of the $\rm{J_{200w}}$ filter at $z=15$ and is luminous enough to impact color. This is tempered by the tendency for this line to become lost against the continuum in brighter galaxies as starbursts comprise of a smaller fraction of the emission. 

For HMXBs, the C IV UV doublet is a constant companion, growing in strength proportionally to the overall luminosity of the halo due to our decision to include them in most of our sample. Since HMXBs can form in metal-enriched populations, diagnostics of this emission line are somewhat extensible to observation of these objects in the local Universe. However, both C IV and the Ca II IR triplet are already a well-established feature of the broad-line regions of nearby accreting compact objects. We find them in our control sample of metal-enriched halos as well which undermines the premise that they are unique to the presence HMXB. We will therefore wait until we perform simulations to lower redshift before we attempt to glean more about the emission-line diagnostics of present-day HMXBs. We also note that our use of solar chemical abundances may significantly underestimate the prevalence of C IV if gas in early galaxies are carbon-enhanced due to the lack of Type Ia supernovae. H-$\alpha$ emission was much more prevalent in our sample of halos with metal-free stars and more luminous in galaxies with a higher fraction of these stars. Though emission is relatively weak compared to the other lines, it may serve as a potential fingerprint for this class of halos.

\section{Conclusion}
\label{sec:con}

We introduce a new radiative transfer post-processing pipeline, {\sc Caius}, for \enzo\ cosmological simulations which we apply to explore the observability of metal-free stellar populations and high mass X-ray binaries. Our main findings are:

\begin{enumerate}
\item High mass X-ray binaries would peak at about 20\% of the stellar systems within a Population III starburst if it is generously assumed that half the stars form as close binaries.
\item About six halos in our sample would be discernible with $JWST$ with long exposure times (1-10 Ms) and gravitational lensing ($\mu = 10$).
\item Galaxies with high fractions of metal-free stars tend to have low luminosity at high redshift. Therefore the best scenario for direct observation of a metal-free stellar population might be a merger between two such galaxies though that configuration is rare in our simulations.
\item The youth of metal-free stars implies strong Ly$\alpha$ emission. Ly$\alpha$ EW are inversely proportional to the total stellar mass of the halo. Through filters, high EW appear as an increase in $\rm{J_{200w} - J_{277w}}$ as compared to their intrinsic values from the underlying stellar spectra.
\item The inclusion of Population III stars and HMXBs significantly increased the prevalence of H-$\alpha$ emission versus the control group and H-$\alpha$ further scaled with the fraction of stellar mass that comes from Population III stars. 
\item Strong Ly$\alpha$ emission gives rise to the Ca II IR triplet, which suffers less extinction than Ly$\alpha$ while indicating the same physical scenario.
\item Our sample of galaxies with Population III stars and HMXB were generally bluer than the control sample.
\end{enumerate}

We have shown the impact of ISM and CGM extinction of the gas and dust continuum as well as emission lines on galaxies with high-energy sources in the early Universe. Our prescription treats extinction and photochemistry in both optically thin and optically thick media. With our pipeline, we are able to produce synthetic photometry and further process those results into instrument-relevant data. We will continue to improve our post-processing models as we explore more cosmological scenarios. 

\section*{Acknowledgements}
KSSB acknowledges support from the Southern Regional Education Board doctoral fellowship.
JHW acknowledges support from National Science Foundation (NSF) grants
AST-1333360 and AST-1614333 and Hubble theory grants HST-AR-13895 and
HST-AR-14326 and NASA grant NNX17AG23G.  
AA acknowledges support fromm NSF grant AST-1333360.
BWO was supported in part by NSF grants PHY-1430152 and AST-1514700, by NASA grants NNX12AC98G, NNX15AP39G, and by Hubble Theory Grants HST-AR-13261.01-A and HST-AR-14315.001-A.
MLN was supported by NSF grant AST-1109243 and acknowledges partial support from NSF grant AST-1615848.    The
simulation was performed using \textsc{Enzo} on the Blue Waters
operated by the National Center for Supercomputing Applications (NCSA)
with PRAC allocation support by the NSF (award number
ACI-0832662). This research is part of the Blue Waters
sustained-petascale computing project, which is supported by the NSF
(award number ACI 1238993 and ACI-1514580) and the state of Illinois. Blue Waters is a
joint effort of the University of Illinois at Urbana-Champaign and its
NCSA.  This research has made use of NASA's Astrophysics Data System
Bibliographic Services.  Analysis was performed on XSEDE's Maverick
resource with XSEDE allocation AST-120046.  The majority of the
analysis and plots were done with \textsc{yt} and \textsc{matplotlib}.
\textsc{Enzo} and \textsc{yt} are developed by a large number of
independent researchers from numerous institutions around the
world. Their commitment to open science has helped make this work
possible.

\bibliography{pop3}
\bibliographystyle{aasjournal}

\bsp
\label{lastpage}

\end{document}